\documentclass[aps,prc,letterpaper,11pt,twoside,tightenlines,nofootinbib,showpacs,preprint]{revtex4-1}
\usepackage{graphicx}
\usepackage[sort&compress]{natbib}
\usepackage{subfigure}
\usepackage{amsmath}
\usepackage{amsfonts}
\usepackage{cancel}
\usepackage{amssymb}
\usepackage{hyperref}
\usepackage{color}
\usepackage{epsfig}

\begin{document}
\arraycolsep1.5pt
\newcommand{\Ima}{\textrm{Im}}
\newcommand{\Rea}{\textrm{Re}}
\newcommand{\mev}{\textrm{ MeV}}
\newcommand{\be}{\begin{equation}}
\newcommand{\ee}{\end{equation}}
\newcommand{\ba}{\begin{eqnarray}}
\newcommand{\ea}{\end{eqnarray}}
\newcommand{\gev}{\textrm{ GeV}}
\newcommand{\nn}{{\nonumber}}
\newcommand{\dtres}{d^{\hspace{0.1mm} 3}\hspace{-0.5mm}}
\newcommand{\rts}{ \sqrt s}
\newcommand{\non}{\nonumber \\[2mm]}
\raggedbottom

\title{Study of $B \bar B^*$ and $B^* \bar{B}^* $ interactions in $I=1$ and relationship to the $Z_b(10610)$, $Z_b(10650)$ states.}

\author{J. M. Dias$^{1,\,2}$, F. Aceti$^{1}$, E. Oset$^{1}$,}
\affiliation{$^{1}$Departamento de F\'{\i}sica Te\'orica, Universidad de Valencia and IFIC, Centro Mixto Universidad de
Valencia-CSIC, Institutos de Investigaci\'on de Paterna, Aptdo. 22085, 46071 Valencia,
Spain\\
\\$^{2}$Instituto de F\'isica, Universidade de S\~ao Paulo,
C.P. 66318, 05389-970 S\~ao Paulo, SP, Brazil\\
 }

\date{\today}

\begin{abstract}
We use the local hidden gauge approach in order to study the $B\bar{B}^*$ and $B^*\bar{B}^*$
interactions for isospin I=1. We show that both interactions via one light meson exchange 
are not allowed by OZI rule and, for that reason, we calculate the contributions due to the 
exchange of two pions, interacting and noninteracting among themselves, and also due to the heavy vector mesons. Then, to compare all these contributions, we use the potential related to the heavy vector exchange as an effective potential corrected by a factor which takes into account the contribution of the others light mesons exchange. In order to look for poles, this effective potential is used as the kernel of the Bethe-Salpeter equation. As a result, for the $B\bar{B}^*$ interaction we find a loosely bound state with mass in the range $10587-10601$ MeV, very close to the experimental value of the $Z_b(10610)$ reported by Belle Collaboration. For the $B^*\bar{B}^*$ case, we find a cusp at $10650$ MeV for all spin $J=0,\,1,\,2$ cases. 
\end{abstract}
\pacs{11.80.Gw, 12.38.Gc, 12.39.Fe, 13.75.Lb} 

\maketitle

\section{Introduction}
\label{Intro}

In 2003 the Belle collaboration observed the first new charmoniumlike state called $X(3872)$ in the $B^{+}\rightarrow X(3872)K^{+}\rightarrow J/\psi\,\pi^+\pi^-K^+$ process \cite{Choi:2003ue}. It was later confirmed by BaBar, CDF and $D0$ collaborations \cite{x3872}. After its discovery, many other new states have been found by these collaborations, with masses situated in the charmonium mass region. Most of them are above the meson-meson threshold and, if they were a conventional charmonium, they would decay into a pair of open charm mesons. However, this is not seen in the experiment, instead, what is observed is their decay into $J/\psi$ plus pions which is an unusual property for a simple $c\bar{c}$ state. Furthermore, the predictions from potential models for the mass and decay channels do not fit with the experimental results. For all these reasons, a strong experimental and theoretical effort has been made in order to understand the quark configuration of these new states as well as their production mechanisms, decay widths, masses and spin-parity assignments. In Refs. \cite{Swanson:2006st,Zhu:2007wz,Nielsen:2009uh,Olsen:2009gi,Brambilla:2010cs,Ali:2011vy} one can find a detailed discussion about the current status of those states, commonly called $X$, $Y$ and $Z$.

Since the discovery of the $X$, $Y$ and $Z$ states, an enormous bulk of work has been done in an attempt to accomodate them in an exotic picture. By exotic we mean a more complex quark structure beyond quark-antiquark state, like hybrid, tetraquark, hadrocharmonium and meson molecule. The exotic state idea is not new, actually is quite old, but before the discovery of $Z^{+}_c(3900)$ by BESIII and Belle collaborations last year, no exotic structure had been conclusively identified. It is a challenge to understand these new charmoniumlike states as exotic since using the models mentioned above it is relatively simple to reproduce the masses of those states. The same challenges also concern the bottomoniumlike states. Among them, the $Z_b(10650)$ and $Z_b(10610)$ are very interesting. They were observed by the Belle collaboration in $\pi^{\pm}\,h_b(nP)$ and $\pi^{\pm}\,\Upsilon(mS)$, with the $n=1, 2$ and $m=1, 2, 3$, invariant mass distribution of the $\Upsilon(5S)$ decay channel \cite{Belle:2011aa}. As a result of the measurements, Belle reported: $M_{Z_b(10610)}=(10608.4\pm 2.0)$ MeV, $\Gamma_{Z_b(10610)}=(15.6\pm 2.5)$ MeV and for $Z_b(10650)$, $M_{Z_b(10650)}=(10653.2\pm 1.5)$ MeV and $\Gamma_{Z_b(10650)}=(14.4\pm 3.2)$ MeV. The quantum numbers are reported as $J^P=1^+$ and positive G parity. The neutral partner has also been observed in the $\Upsilon(5S) \to \Upsilon(nS) \pi \pi$ decay in the belle Colaboration \cite{Adachi:2012im}.

In an attempt to understand the $Z_b(10610)$ and $Z_b(10650)$ configuration, some interpretations were considered. The authors of \cite{Bondar:2011ev} treated the states as molecular states of  $B \bar B^*$ and $B^* \bar B^*$ using HQSS, but the strength of the interaction was unknown. The proximity of the masses of these states to the $B \bar B^*$ and $B^* \bar B^*$ thresholds prompted the author of \cite{Bugg:2011jr} to suggest that these peaks could be a consequence of cusps originated at these thresholds. This idea has been made more quantitative in a recent paper \cite{Swanson:2014tra}. In \cite{Danilkin:2011sh} the dynamics of hadro-quarkonium system was formulated, based on the channel coupling of a light hadron (h) and heavy quarkonium ($Q\,\bar Q$) to intermediate open-flavor heavy-light mesons (Qq,  Qq). In \cite{Cui:2011fj} the authors used QCD sum rules assuming tetraquarks or molecules, and in all cases they could obtain good results, but the errors in the masses were of the order of 200 to 300 MeV. In the same line, in \cite{Guo:2011gu} the states are also assumed to be tetraquarks. A tetraquark picture was assumed by the authors of Ref. \cite{Chen:2011}, where using the framework of QCD sum rules, they calculated the $Z_b$'s mass, but the masses obtained were lower than those of the $Z_b$ states. In \cite{Sun:2011uh} the authors consider the states as molecular states driven by the one pion exchange interaction. In \cite{Cleven:2011gp} heavy quark spin symmetry is used, analysing the power counting of the loops, and concluding that the molecular nature of the states can account for the observed features. In  \cite{Voloshin:2011qa} the authors mention that using heavy quark spin symmetry (HQSS) and the molecular picture, states of $1^-$ should exist in addition to the reported states of $1^+$. In \cite{Zhang:2011jja} the molecular option is also supported by sum rules, but again with about 220 MeV uncertainty in the mass. In \cite{Ali:2011ug} a tetraquark nature is invoked. 
In  \cite{Yang:2011rp} $B \bar B^*, ~ B^* \bar B^*$  (in S-wave) are investigated in the framework of chiral quark models using the Gaussian expansion method. The bound states of $B \bar B^*, ~ B^* \bar B^*$ with quantum numbers $I(J^{P}) = 1(1^{+})$, which are good candidates for the $Z_b(10610)$ and $Z_b(10650)$ respectively, are obtained. Another $B \bar B^*$ bound state with $I(J^{PC}) = 0(1^{++})$, and other two  $B^* \bar B^*$ with $I(J^{PC}) = 1(0^{++})$, $I(J^{PC}) = 0(2^{++})$ are predicted in that work. In \cite{Guo:2011gu} $1^+$ tetraquarks are invoked and possible $1^{++}$ , $2^{++}$ states from charge conjugation are investigated. In  \cite{Chen:2011zv} the molecular picture is again pursued and the $\Upsilon(5S) \to  \Upsilon(nS) \pi^+ \pi^-$ decays are investigated. In \cite{Nieves:2011zz} the authors make arguments of HQSS starting from the X(3872) extrapolating to the beauty sector, and find a plausible molecular interpretation for the $Z_b(10610)$ state. In \cite{Mehen:2011yh} once again the molecular structure is supported within HQSS. A different intepretation is given in  \cite{Chen:2011pv}, where the initial pion emission mechanism is invoked to reproduce the $\Upsilon(5S) \to \Upsilon(nS) \pi \pi$, with the second $\pi$ and the resonance produced from the loop diagram involving three $B^*$ states. 
Again from the molecular point of view in \cite{Cleven:2013sq}, several decay channels are investigated in order to give support for the molecular picture. In  \cite{Dong:2012hc}, using phenomenological Lagrangians and the hypothesis of molecular states, the $Z \to \Upsilon(nS) \pi$ transition rates are evaluated. Tetraquarks are again invoked in \cite{Braaten:2013boa}.
Pion exchange is considered in  \cite{Valderrama:2012jv}  and limits for the strength to produce binding are discussed.  In \cite{Ke:2012gm} a tetraquark is preferred, since meson exchange binds in I=0 but not in I=1.  By using HQSS and assuming the states to be molecular states, different modes of production are evaluated in \cite{Ohkoda:2012rj}.

Using the chiral quark models, the authors of \cite{Li:2012wf} interpret the states as loosely bound states of $B \bar B^*, ~ B^* \bar B^*$. Tetraquarks are again favoured in sum rules in \cite{Wang:2013zra}. In \cite{Guo:2013gka} the authors use HQSS to relate these states, which are assumed to be molecular, to the X(3872). A molecular interpretation was again used in Ref. \cite{Zhi:2011liu} in order to explain the states as $B^*\bar{B}$ and $B^*\bar{B}^*$ assuming a s- and d-wave mixture.

The amount of theoretical work done is quite large, offering theoreticians a challenge with observed states that obviously cannot have a $c\bar{c}$ nature, which should have $I=0$. Our contribution to the subject lines up with the molecular interpretation, using a dynamical model that provides the strength of the interaction. We use for this purpose the extrapolation of the local hidden gauge approach to the heavy sector, extending results obtained for the $Z_c(3900)$ and $Z_c(4025)$ using that approach \cite{Aceti:2014kja,Aceti:2014uea} which at the same time was shown to fully respect the rules of HQSS \cite{xiaoliang,xiaojuan}.

\section{Formalism}
\label{Formalism}

\begin{figure}[htpb]
\centering
\includegraphics[scale=0.65]{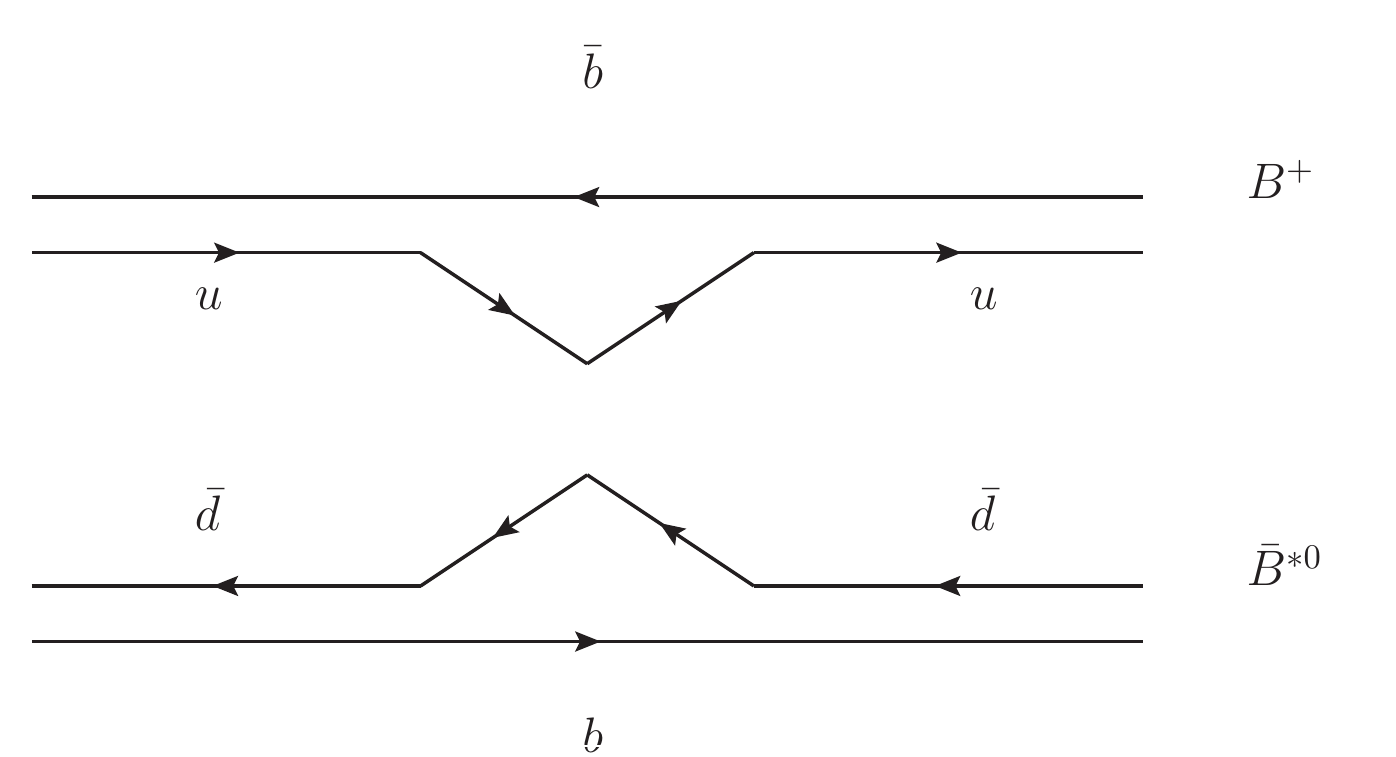}
\caption{Diagram representing the $B^+\bar{B}^{*0}\rightarrow B^+\bar{B}^{*0}$ process through the exchange of $q\bar{q}$, which is not allowed by OZI rule.}
\label{fig:OZI}
\end{figure}

In order to study $B\bar{B}^*$ and $B^*\bar{B}^*$ states, the extension of the local hidden gauge approach \cite{hidden1,hidden2,hidden3} to the heavy quark sector \cite{raquelxyz} seems most appropriate. The interaction is generated by the exchange of a vector meson. If one exchanges light vectors the heavy quarks act as spectators and then, the heavy quark spin symmetry (HQSS) of QCD is automatically fulfilled \cite{xiaoliang}. However, following the approach of Refs. \cite{Aceti:2014kja,Aceti:2014uea}, we can show that the $B\bar{B}^*$ and $B^*\bar{B}^*$ interactions by means one light meson exchange are not allowed by OZI rule for $I=1$ states. In Fig. \ref{fig:OZI}, a diagram illustrating an interaction between a $B^+\bar{B}^{*0}$ is shown. In order for this interaction to occur a $d\bar{d}$ state has to be converted into a $u\bar{u}$ state, which is OZI forbidden. This implies a cancellation between the contributions coming from $\rho$ and $\omega$ mesons exchange if equal masses are taken. The same argument holds for the exchange of a pseudoscalar and one finds an exact cancellation of $\pi,\,\eta,\, \eta^{\prime}$ exchange in the limit of equal masses for these mesons \cite{Aceti:2014kja,Aceti:2014uea}.

Because of this cancellation, we shall consider processes in which the OZI restriction no longer holds. We, thus, calculate the contributions coming from heavy vector exchange and also due to the exchange of two pions, interacting and non-interacting among themselves. 


\subsection{$B\bar{B}^*$ and $B^*\bar{B}^*$ interactions via heavy vector exchange}
\label{vecex}


In order to evaluate $B\bar{B}^*$ and $B^*\bar{B}^*$ interactions due to the exchange of vector mesons, we need the Lagrangians describing the $VPP$ and $VVV$ vertices, namely
\begin{equation}
\mathcal{L}_{VPP}=-ig\langle V^{\mu}[P,\partial_{\mu}P]\rangle\ ,
\label{eq:VPPlag}
\end{equation}
\begin{equation}
\mathcal{L}_{VVV}=ig\langle (V^{\mu}\partial_{\nu}V_{\mu}-\partial_{\nu}V_{\mu}V^{\mu})V^{\nu}\rangle\ 
\label{eq:VVVlag}.
\end{equation}
The coupling $g$ is given by $g=M_{V}/2f_{\pi}$, being $f_{\pi}=93$ MeV the pion decay constant, while $M_V$ is the vector meson mass.

In Eqs. \eqref{eq:VPPlag} and \eqref{eq:VVVlag}, the symbol $\langle \, \rangle$ stands for the trace of SU(4). The vector field $V_{\mu}$ is represented by the SU(4) matrix, which is parametrized by 16 vector mesons including the 15-plet and singlet of SU(4), 
\begin{equation}
V_\mu=\left(
\begin{array}{cccc}
\frac{\omega}{\sqrt{2}}+\frac{\rho^0}{\sqrt{2}} & \rho^+ & K^{*+}&\bar{B}^{*0}\\
\rho^- &\frac{\omega}{\sqrt{2}}-\frac{\rho^0}{\sqrt{2}} & K^{*0}&B^{*-}\\
K^{*-} & \bar{K}^{*0} &\phi&B^{*-}_s\\
B^{*0}&B^{*+}&B^{*+}_s&J/\psi
\end{array}
\right)_\mu\ ,
\label{eq:vfields}
\end{equation}
where the ideal mixing has been taken for $\omega$, $\phi$ and $J/\psi$. On the other hand, $P$ is a matrix containing the 15-plet of the pseudoscalar mesons written in the physical basis in which $\eta$, $\eta^{\prime}$ mixing is taken into account \cite{gamphi3770},
\begin{equation}
P=\left(
\begin{array}{cccc}
\frac{\eta}{\sqrt{3}}+\frac{\eta'}{\sqrt{6}}+\frac{\pi^0}{\sqrt{2}} & \pi^+ & K^+&\bar{B}^0\\
\pi^- &\frac{\eta}{\sqrt{3}}+\frac{\eta'}{\sqrt{6}}-\frac{\pi^0}{\sqrt{2}} & K^{0}&B^-\\
K^{-} & \bar{K}^{0} &-\frac{\eta}{\sqrt{3}}+\sqrt{\frac{2}{3}}\eta'&B^-_s\\
B^0&B^+&B^+_s&\eta_b
\end{array}
\right)\ .
\label{eq:pfields}
\end{equation}

The channels we are interested in are those with $B=0$, $S=0$ and isospin $I=1$. In the $B^*\bar{B}^*$ case, they are $B^*\bar{B}^*$ and $\rho\Upsilon$. In the case of $B\bar{B}^*$ we are only interested in the positive $G$-parity combination, namely $(B\bar{B}^*+cc)/\sqrt{2}$ and also $\eta_{b}\,\rho$ and $\pi\,\Upsilon$.


\subsubsection{$B^*\bar{B}^*$ case}


\begin{figure}[htpb]
\centering
\includegraphics[scale=0.65]{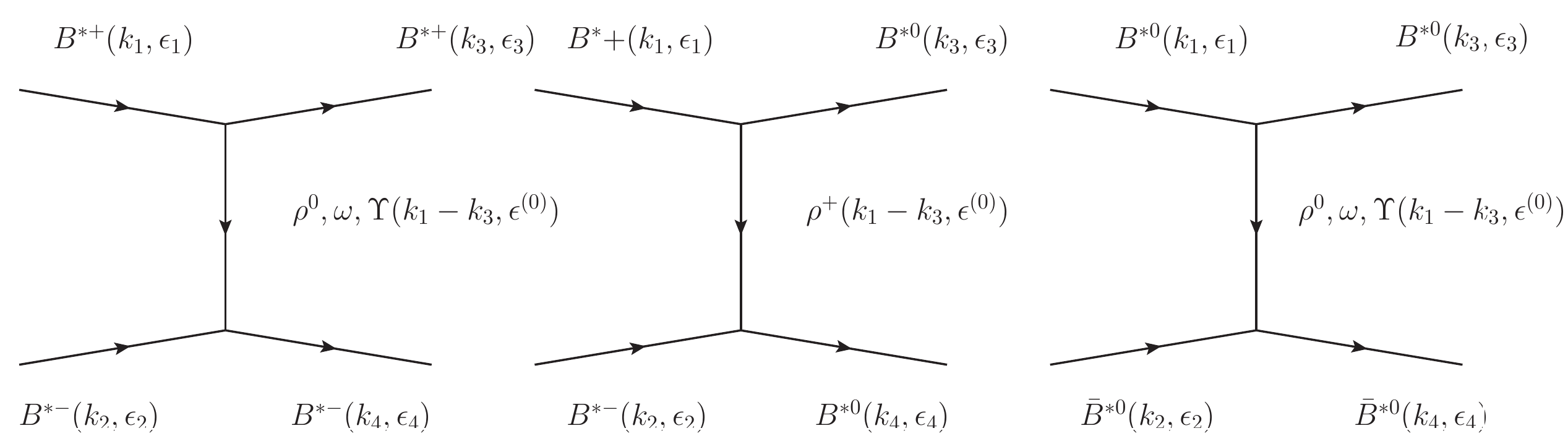}
\caption{Vector exchange diagrams contributing to the process $B^*\bar{B}^*\rightarrow B^*\bar{B}^*$.}
\label{fig:vecdiag}
\end{figure}

\begin{figure}[htpb]
\centering
\includegraphics[scale=0.65]{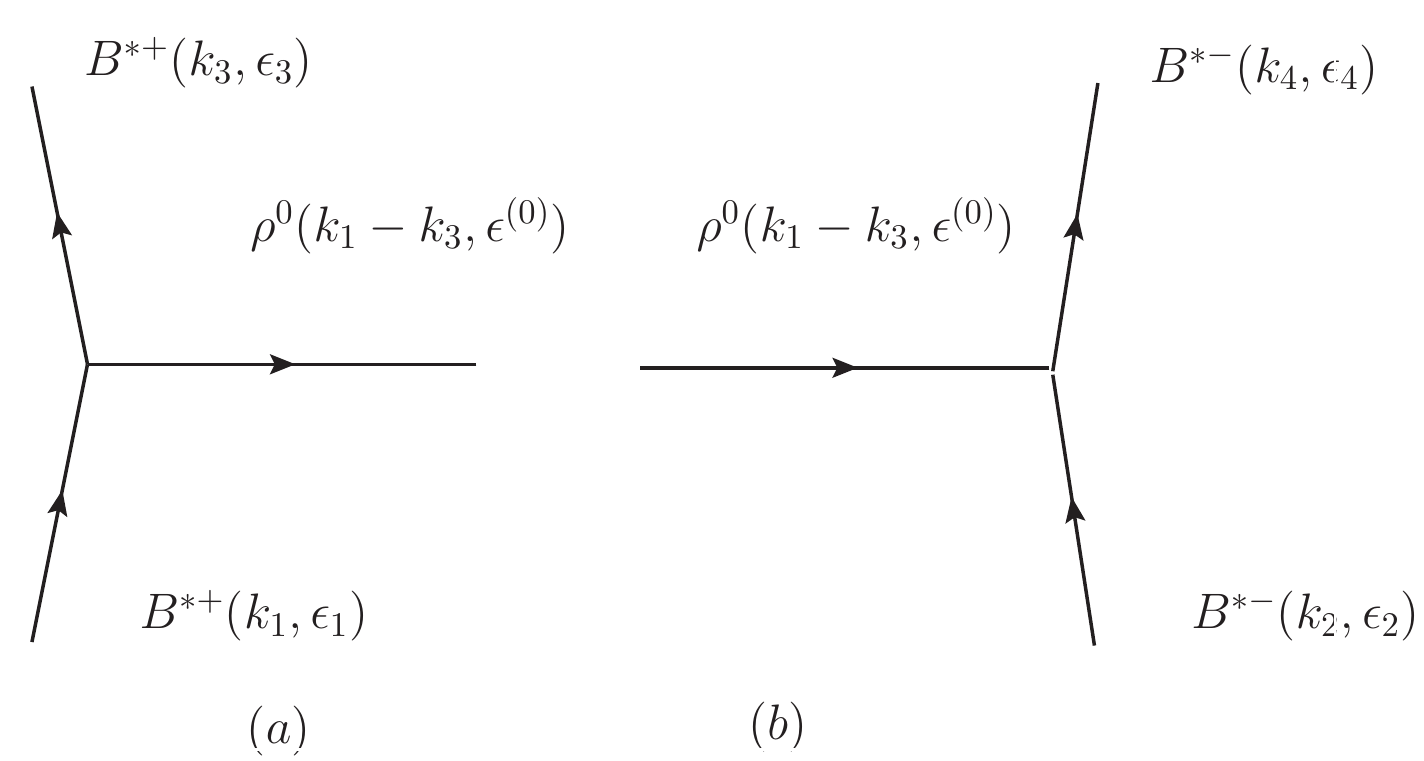}
\caption{Three-vector vertex associated with $B^{*+}\,B^{*+}\,\rho^0$.}
\label{fig:vecvertex}
\end{figure}

Consider now the reaction $B^*\bar{B}^*\rightarrow B^*\bar{B}^*$. Here we are following the same steps as in Ref. \cite{raquelxyz}, in which the authors were concerned in the $D^*\bar{D}^*$ case. As in \cite{raquelxyz} we also consider that the external vectors have negligible three-momentum with respect to their masses. In our case, the most important diagrams are depicted in Fig. \ref{fig:vecdiag}. As an example, we shall calculate in detail the amplitude of the first diagram in Fig. \ref{fig:vecdiag}. The evaluation of the other ones is analogous. For this end, we must calculate the three-vector vertex which is given by the Lagrangian of Eq. \eqref{eq:VVVlag}. Figs. \ref{fig:vecvertex}(a) and (b) illustrate the three-vector vertices $B^{*+}\bar{B}^{*+}\rho^{0}$ and $B^{*-}\bar{B}^{*-}\rho^{0}$ with the momenta assignments. The corresponding vertex functions are
\begin{equation}
t_{B^{*+}B^{*+}\rho^0}=\frac{g}{\sqrt{2}}(k_1+k_3)_{\mu} \epsilon_{1\nu}\epsilon_{3}^{\nu} \epsilon_{\mu}^{(0)},
\end{equation}
\begin{equation}
t_{B^{*-}B^{*-}\rho^0}=\frac{g}{\sqrt{2}}(k_2+k_4)_{\mu} \epsilon_{2\nu}\epsilon_{4}^{\nu} \epsilon_{\mu}^{(0)}.
\end{equation}

Once we have determined the vertices, it is possible to calculate the amplitude for the first diagram of Fig. \ref{fig:vecdiag}. Considering all the particles involved in the exchange, we obtain
\begin{eqnarray}
t_{B^{*+}B^{*-}\rightarrow B^{*+}B^{*-}}=-\frac{1}{2}g^2\left[\frac{2}{M_{\Upsilon}^2}+\frac{1}{M_{\rho}^2}+\frac{1}{M_{\omega}^2}\right](k_1+k_3)\cdot (k_2+k_4)\epsilon_{1\mu}\epsilon_{2\nu}\epsilon_3^{\mu}\epsilon_4^{\nu}\, ,
\label{eq:tbb}
\end{eqnarray}
where $M_{\Upsilon}$, $M_{\rho}$ and $M_{\omega}$ are the masses of the $\Upsilon$, $\rho$ and $\omega$ mesons, respectively.

As we are interested in the $B^*\bar{B}^*$ interaction in the $I=1$ channel, we must rewrite Eq. \eqref{eq:tbb} in the isospin basis. The isospin states are
\begin{eqnarray}
|B^*\bar{B}^*\rangle^{I=1} = -\frac{1}{\sqrt{2}}|B^{*+}\bar{B}^{*-}\rangle+\frac{1}{\sqrt{2}}|B^{*0}\bar{B}^{*0}\rangle\, ,\\ \nonumber
|B^*\bar{B}^*\rangle^{I=0} = \frac{1}{\sqrt{2}}|B^{*+}\bar{B}^{*-}\rangle-\frac{1}{\sqrt{2}}|B^{*0}\bar{B}^{*0}\rangle\, .
\end{eqnarray}
By taking into account all the three diagrams of Fig. \ref{fig:vecdiag}, we get 
\begin{eqnarray}
t^{I=1}_{B^*\bar{B}^*\rightarrow B^*\bar{B}^*}=g^2\left[\frac{2M_{\rho}^2 M_{\omega}^2+M_{\Upsilon}^2(-M^{2}_{\omega}+M^2_{\rho})}{2M_{\Upsilon}^2M^{2}_{\omega}M_{\omega}^2}\right](k_1+k_3)\cdot (k_2+k_4)\epsilon_{1\mu}\epsilon_{2\nu}\epsilon_3^{\mu}\epsilon_4^{\nu}
\label{eq:tbbI1},
\end{eqnarray}
which shows explicitly the cancellation of $\rho$ and $\omega$ exchange.

In order to rewrite the amplitude given by Eq. \eqref{eq:tbbI1} in terms of spin $0$, $1$ and $2$ states, we use the spin projectors $\mathcal{P}^{(0)}$, $\mathcal{P}^{(1)}$ and $\mathcal{P}^{(2)}$ given by \cite{raquelxyz}
\begin{equation}
\begin{split}
&\mathcal{P}^{(0)}=\frac{1}{3}\epsilon_{\mu}\epsilon_{\mu}\epsilon^{\nu}\epsilon^{\nu}\ ,\\
&\mathcal{P}^{(1)}=\frac{1}{2}\left(\epsilon_{\mu}\epsilon_{\nu}\epsilon^{\mu}\epsilon^{\nu}-\epsilon_{\mu}\epsilon_{\nu}\epsilon^{\nu}\epsilon^{\mu}\right)\ ,\\
&\mathcal{P}^{(2)}=\frac{1}{2}\left(\epsilon_{\mu}\epsilon_{\nu}\epsilon^{\mu}\epsilon^{\nu}+\epsilon_{\mu}\epsilon_{\nu}\epsilon^{\nu}
\epsilon^{\mu}\right)-\frac{1}{3}\epsilon_{\mu}\epsilon_{\mu}\epsilon^{\nu}\epsilon^{\mu}\ ,
\label{eq:projection}
\end{split}
\end{equation}
where the order of the particles $1$, $2$, $3$ and $4$ is implicit. In terms of those projectors the polarization vector combination $\epsilon_{1\mu}\epsilon_{2\nu}\epsilon_{3}^{\mu}\epsilon_{4}^{\nu}$ appearing in Eq. \eqref{eq:tbbI1} is equal to
\begin{equation}
\epsilon_{1\mu}\epsilon_{2\nu}\epsilon_{3}^{\mu}\epsilon_{4}^{\nu}=\mathcal{P}^{(0)}+\mathcal{P}^{(1)}+ \mathcal{P}^{(2)}.
\label{eq:epprojec}
\end{equation}

Therefore, substituting Eq. \eqref{eq:epprojec} into Eq. \eqref{eq:tbbI1}, projecting it in s-wave, and including the contact term already evaluated in Ref. \cite{raquelxyz}, we obtain
\begin{eqnarray}
t^{I=1,S=0,1,2}_{B^*\bar{B}^*\rightarrow B^*\bar{B}^*}=-g^2+g^2\left[\frac{2M_{\rho}^2 M_{\omega}^2+M_{\Upsilon}^2(-M^{2}_{\omega}+M^2_{\rho})}{4M_{\Upsilon}^2M^{2}_{\omega}M_{\rho}^2}\right](4M_{B^*}^2-3s)\, ,
\label{eq:tbb2}
\end{eqnarray}
where $s$ stands for the center of mass energy of the $B^*\bar{B}^*$ system.

\begin{figure}[htpb]
\centering
\includegraphics[scale=0.6]{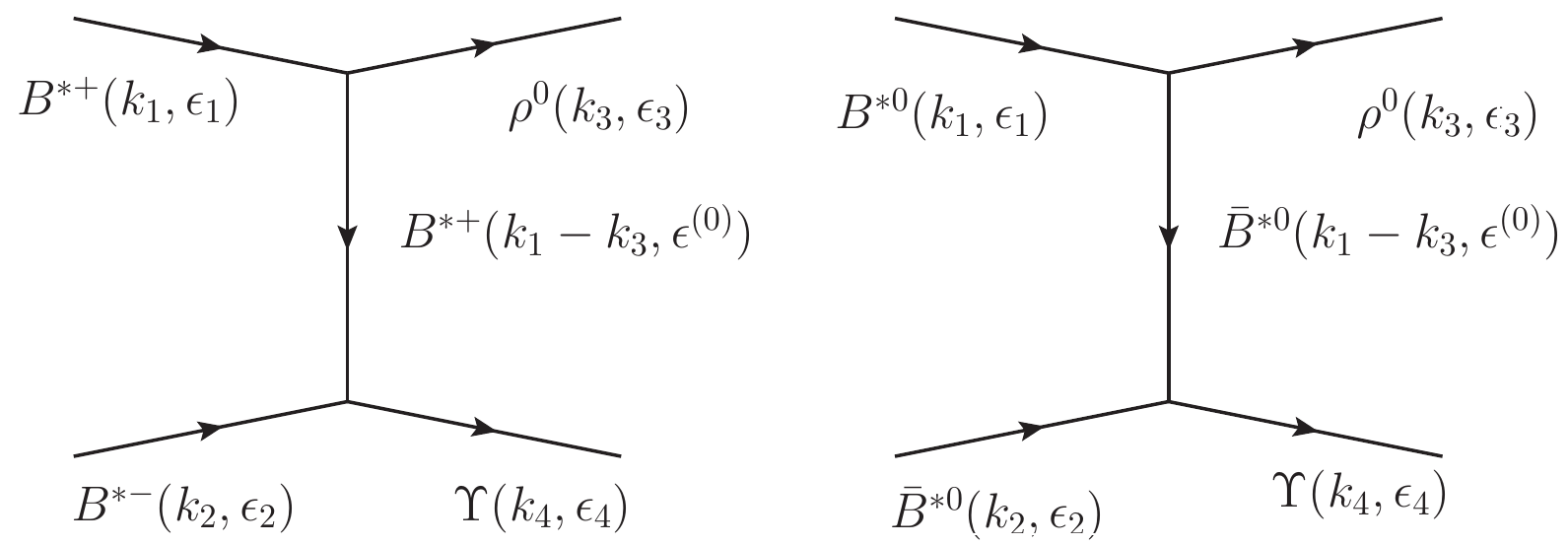}
\caption{Vector exchange diagrams contributing for the $B^*\bar{B}^*\rightarrow \rho\,\Upsilon$ channel.}
\label{fig:vecBdiag}
\end{figure}

Consider now the other channel, $B^*\bar{B}^*\rightarrow \rho \Upsilon$. The most relevant diagrams are depicted in Fig \ref{fig:vecBdiag}. The procedure to get the amplitude for this channel is analogous to what we have done earlier. Thus, the amplitude in isospin $I=1$ basis for the spin $S=0,\,2$ states in s-wave, corresponding to all diagrams of Fig. \ref{fig:vecBdiag} plus the contact term is given by
\begin{equation}
t_{B^*\bar{B}^*\rightarrow \rho\, \Upsilon}^{I=1,S=0,2}=-2g^2+g^2 \left[\frac{2M^2_{B^{*}}+M^2_{\Upsilon}+M^2_{\rho}-3s}{M^2_{B^*}}\right].
\label{eq:trhoupsilon}
\end{equation}
The interaction in $S=1$ vanishes as a consequence of a cancellation of terms where the $\rho^0$ and $\Upsilon$ are interchanged in the diagrams. The diagonal $\rho\Upsilon \to \rho \Upsilon$ transition is again OZI forbidden and null in this approach.

Eqs. \eqref{eq:tbb2} and \eqref{eq:trhoupsilon} will be used as a kernel of the Bethe-Salpeter equation as we shall discuss it later.


\subsubsection{$B\bar{B}^*$ case}


In this case, the Lagrangians defined in Eqs. \eqref{eq:VPPlag} and \eqref{eq:VVVlag} can also be used to provide the vertices of the $PV\rightarrow PV$ interaction through exchange of a heavy vector. The resulting amplitudes were already calculated in s-wave in Refs. \cite{luisaxial,daniel}. In particular the authors were concerned with axial-vector resonances dynamically generated. Yet, in Ref. \cite{Aceti:2014uea} the same equation for the amplitude is used in order to study $D\bar{D}^*$ interaction. Here, we extend these amplitudes for the $B\bar{B}^*$ interaction in the isospin $I=1$ channel, with the result
\begin{equation}
V_{ij}(s)=-\frac{\vec{\epsilon}\ \vec{\epsilon}\ '}{8f_{\pi}^2}\,\mathcal{C}_{ij}\left[3s-(M^2+m^2+M'^2+m'^2)- \frac{1}{s}(M^2-m^2)(M'^2-m'^2)\right]\ ,
\label{eq:VVPPamp}
\end{equation}
where the masses $M$ ($M'$) and $m$ ($m'$) in Eq. \eqref{eq:VVPPamp} correspond to the initial (final) vector meson and pseudoscalar meson, respectively. The indices $i$ and $j$ represent the initial and final $VP$ channels $(B\bar{B}^*+cc)/\sqrt{2}$, $\eta_{b}\,\rho$ and $\pi\,\Upsilon$.

The $\mathcal{C}_{ij}$ are elements of a $3\times 3$ matrix, which for the positive $G$-parity of the $B\bar{B}^*$ combination, is defined as
\begin{equation}
\label{eq:cmatrix}
\mathcal{C}_{ij}=\left( \begin{array}{ccc}
-\psi\ \ \  & \sqrt{2}\gamma\ \ \  & \sqrt{2}\gamma \\
\sqrt{2}\gamma\ \ \  & 0\ \ \  & 0 \\
\sqrt{2}\gamma\ \ \  & 0\ \ \  & 0 \end{array} \right)\ ,
\end{equation}
where $\gamma=\left(\frac{m_L}{m_H}\right)^2$ and $\psi=\left(\frac{m_L}{m_{H\prime}}\right)^2$. Those factors are defined in this way in order to take into account the suppression due to the exchange of a heavy vector meson. Concerning the parameters $m_L$, $m_H$ and $m_{H\prime}$, we choose their values in order to have the same order of magnitude of the light and heavy vector meson masses: $m_L=800$, $m_H=5000$ MeV and $m_{H\prime}=9000$ MeV.



\subsubsection{The $T$-Matrix}
\label{tmatrix}

The results of the amplitudes discussed earlier provide the potential or kernel to be used in the Bethe-Salpeter equation in coupled channels,
\begin{equation}
T=(1-VG)^{-1}V\ ,
\label{eq:bs}
\end{equation}
where $V$ is the potential, which in the $B^*\bar{B}^*$ case is a $2\times 2$ matrix whose elements are the amplitudes defined by Eqs. \eqref{eq:tbb2} and \eqref{eq:trhoupsilon} respectively associated with the channels $B^*\bar{B}^*$ and $\rho\, \Upsilon$. In the case of $B\bar{B}^*$, $V$ is a $3\times 3$ matrix and its elements are the amplitudes given by Eq. \eqref{eq:VVPPamp} with $\mathcal{C}_{ij}$ defined by Eq. \eqref{eq:cmatrix}, associated with the channels $B\bar{B}^*$, $\eta_{b}\, \rho$ and $\pi\, \Upsilon$.

In Eq. \eqref{eq:bs}, $G$ is a diagonal matrix and its elements are given by the two meson loop function, $G_l$ for each channel $l$:
\begin{equation}
G_l=i\int\frac{d^4q}{(2\pi)^4}\frac{1}{q^2-m^2+i\epsilon}\frac{1}{(q-P)^2-M^2+i\epsilon}\ ,
\label{eq:loopex}
\end{equation}
where $m$ is the mass of the pseudoscalar (in the $B\bar{B}^*$ case) or vector (in the $B^*\bar{B}^*$ case), while $M$ is the vector meson mass involved in the loop in the channel $l$. In Eq. \eqref{eq:loopex} $P$ means the total four-momentum of the mesons. The integral of Eq. \eqref{eq:loopex} is logarithmically divergent and it can be regularized with a cut off in the momentum space or dimensional regularization. With the cut off method
\begin{equation}
\label{eq:1striemann}
G_l=\int\limits_{0}^{q_{max}}\frac{d^3q}{(2\pi)^3}\,\frac{\omega_1+\omega_2}{2\omega_1\omega_2}\,\frac{1}{(P^0)^2-(\omega_1+\omega_2)^2+i\epsilon}\ ,
\end{equation}
where $\omega_1=\sqrt{m^2+\vec{q}^{\ 2}}$ and $\omega_2=\sqrt{M^2+\vec{q}^{\ 2}}$ and $q_{max}$ is a free parameter. In dimensional regularization there is a scale $\mu$ and a subtraction constant $\alpha(\mu)$ acting as a free parameter, namely,
\begin{equation}
\begin{split}
\label{eq:loopexdm}
G_l&=\frac{1}{16\pi^2}(\alpha_l+\log\frac{m^2}{\mu^2}+\frac{M^2-m^2+s}{2s}\log\frac{M^2}{m^2}+\frac{p}{\sqrt{s}}(\log\frac{s-M^2+m^2+2p\sqrt{s}}{-s+M^2-m^2+2p\sqrt{s}}\\&+\log\frac{s+M^2-m^2+2p\sqrt{s}}{-s-M^2+m^2+2p\sqrt{s}}))\ .
\end{split}
\end{equation}
with $p$ standing for the three-momentum of the mesons in the center-of-mass frame.

For the sake of comparison of the different potentials obtained, it is interesting to recall that Eq. \eqref{eq:bs} with the cut off regularization of Eq. \eqref{eq:1striemann} can be obtained from the Lippmann-Schwinger equation using a potential in momentum space \cite{daniwf}
\begin{equation}
V(\vec{q},\vec{q\,}^{\prime})=V\theta(q_{max}-|\vec{q\,}|)\theta(q_{max}-|\vec{q\,}^{\prime}|)\, .
\label{eq:potheta}
\end{equation}

Hence, assuming $\vec{q}\approx 0$ for an external particle, $\vec{q\,}^{\prime}$ can play the role of momentum transfer in loop diagrams, and then $V$ as a function of $\vec{q\,}^{\prime}$ remains constant up to $q_{max}$, where it goes to zero.


\subsection{The $\sigma$ exchange contribution to the $B\bar{B}^*$ and $B^*\bar{B}^*$ interactions}


The potential due to the $\sigma$ exchange in some cases provides an important contribution to the interaction. In Ref. \cite{toki} the authors studied the $NN$ system considering that the $\sigma$ resonance arises from the interaction of two pions, providing an important contribution to the binding energy for the $NN$ system. In Refs. \cite{Aceti:2014uea,Aceti:2014kja} the same idea was applied to the $D\bar{D}^*$ and $D^*\bar{D}^*$ cases. Following the approach of those references we shall extend the formalism to the bottom sector, more specifically, to study the $B\bar{B}^*$ and $B^*\bar{B}^*$ interactions.

\begin{figure}[htpb]
\centering
\includegraphics[scale=0.5]{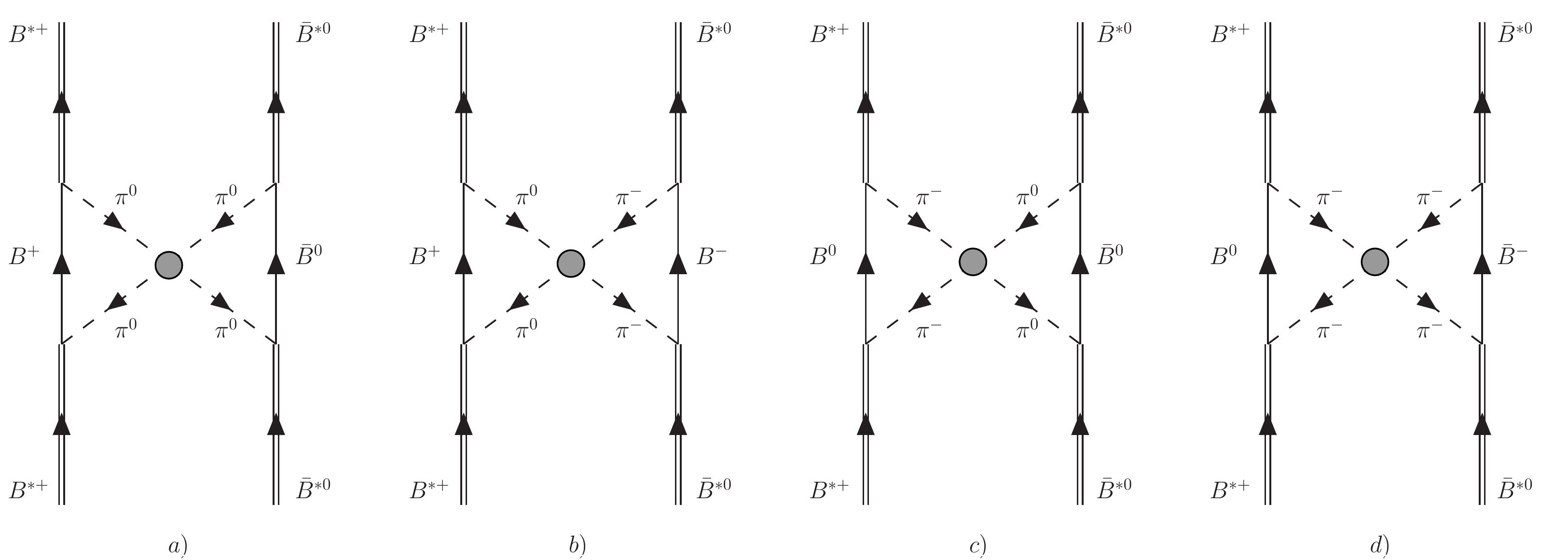}
\caption{Diagrams contributing to the two pions interaction in lowest order in $I=1$ for the $B^*\bar{B}^*\rightarrow B^*\bar{B}^*$ process.}
\label{fig:diagsigma}
\end{figure}

Let us consider first, the $B^*\bar{B}^*$ case. The diagrams contributing to this interaction are illustrated in Fig. \ref{fig:diagsigma}. As can be seen from Fig. \ref{fig:diagsigma}, each diagram has four vertices containing two pseudoscalars, the $\pi$ and $B\,(\bar{B})$ mesons and one $B^*\,(\bar{B}^*)$ vector. Their evaluation is done by means of the local hidden gauge Lagrangian already defined in Eq. \eqref{eq:VPPlag}. On the other hand, instead of calculating the vertices and then the amplitude from the Lagrangian of Eq. \eqref{eq:VPPlag}, we start from the amplitude obtained in Ref. \cite{Aceti:2014kja} and substitute the masses of the $D$ and $D^*$ mesons by the masses of the $B$ and $B^*$, respectively. As a result, we obtain 
\begin{equation}
-it_{B^*\bar{B}^*}^{\sigma}=-i\ V^2\ \frac{3}{2}\ t_{\pi\pi\rightarrow\pi\pi}^{I=0}\ ,
\label{eq:ampl2}
\end{equation}
where $t^{I=0}_{\pi\,\pi\rightarrow \pi\,\pi}$ is the isoscalar amplitude for the $\pi\,\pi$ interaction, namely
\begin{equation}
t^{I=0}_{\pi\,\pi\rightarrow \pi\,\pi}=-\frac{1}{f}\frac{s^{\prime}-\frac{m^2_{\pi}}{2}}{1+\frac{1}{f^2}G(s^{\prime})(s^{\prime}-\frac{m^2_{\pi}}{2})}\, ,
\label{eq:isoampl}
\end{equation}
with $G(s^{\prime})$ the two pions loop function suited to this case (whose explicit form is given in \cite{toki,Aceti:2014kja}) and with $P$ the total $\pi\,\pi$ momentum, with the pions travelling to the right in the diagrams. Hence, $P^2=s^{\prime}$ is actually the variable $t$ for the $B^{*+}\bar{B}^{*0}$ system. 


In Eq. \eqref{eq:ampl2}, $V$ is a factor that takes into account the contributions coming from the triangular loops of the diagram. The detailed derivation of the $V$ factor can be found in Ref. \cite{Aceti:2014kja}. We adopt the Breit frame, 
\begin{equation}
\begin{split}
&p_1\equiv(p_1^0, \vec{q}/2)\ ,\\
&p_1'\equiv(p_1'^{\ 0}, -\vec{q}/2)\ ,\\
&p\equiv(p^0, \vec{p}\,)\ ,
\label{eq:breit}
\end{split}
\end{equation}
where $\vec{q}$ is the three-momentum transferred in the process and $p_1$ and $p^{\prime}_1$ the momenta for the two incoming $B^*$. The equations for $V$ are obtained from \cite{Aceti:2014kja} with the trivial changes in the masses of the particles. It also contains the factor $(M_{B^*}/M_{K^*})^4$ replacing the factor $(M_{D^*}/M_{K^*})^4$ in \cite{Aceti:2014kja} as demanded by HQSS in \cite{xiaoliang}.

\begin{figure}[htpb]
\centering
\includegraphics[scale=0.8]{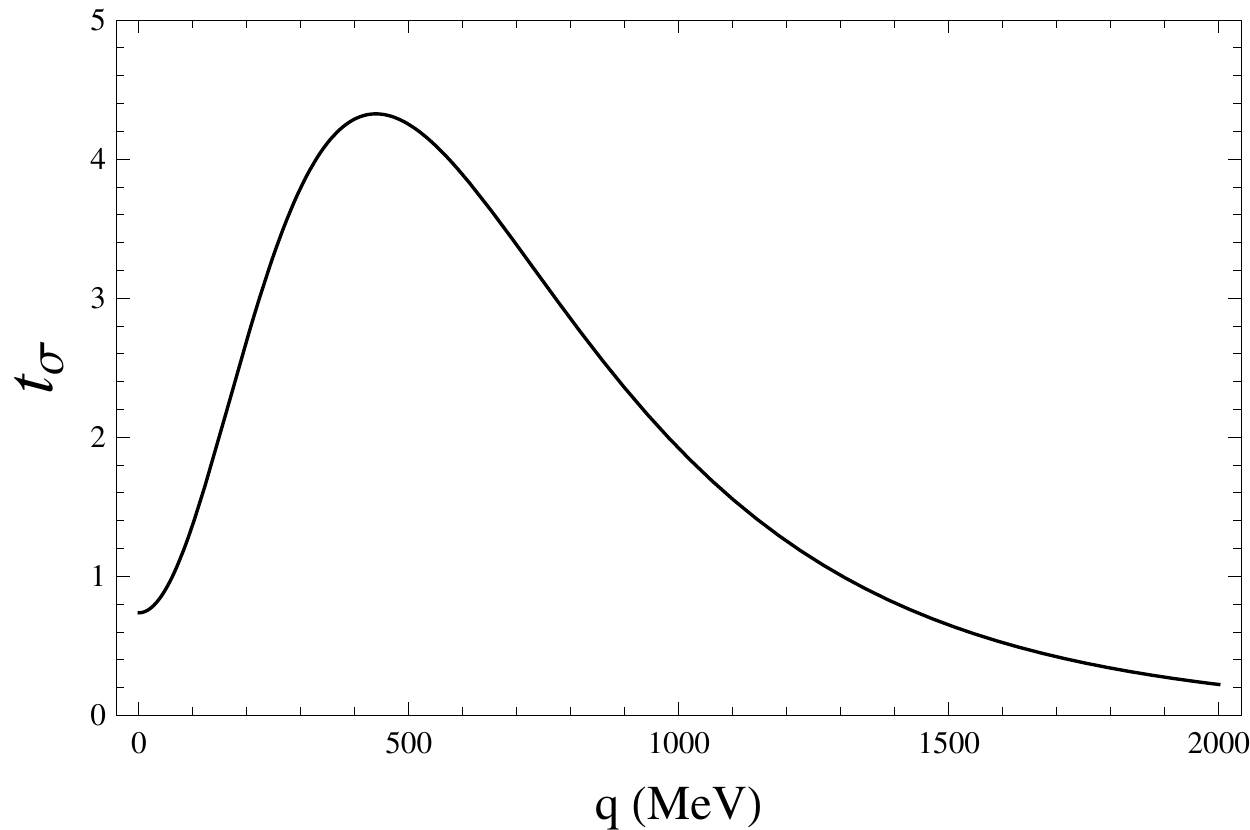}
\caption{Potential $t^{\sigma}_{B^*\bar{B}^*}$ as a function of the momentum transferred in the process.}
\label{fig:tsigma}
\end{figure}

Finally, substituting Eq. \eqref{eq:isoampl} into Eq. \eqref{eq:ampl2} and taking $s=-\vec{q\,}^2$ since there is no energy exchange, we get the following expression for the potential
\begin{equation}
t_{B^*\bar{B}^*}^{\sigma}(\vec{q})=V^2\,\frac{3}{2}\,\frac{1}{f^2}\,\frac{\vec{q}^{\,2}+\frac{m_{\pi}^2}{2}}{1-G(-\vec{q}^{\,2})\,\frac{1}{f^2}(\vec{q}^{\,2}+\frac{m_{\pi}^2}{2})}\ ,
\label{eq:ampl-fin}
\end{equation}
with 
\begin{equation}
V=\epsilon_{\mu}\epsilon'_{\nu}(ag^{\mu\nu}+cp_1'^{\mu}p_1^{\nu})
\label{eq:V3}
\end{equation}
and $a$ and $c$ also given in \cite{Aceti:2014kja} with trivial changes in the masses. Assuming the spatial components of the momenta $p_{1\mu}$ and $p_{1\nu}^{\prime}$ smaller than the vector masses, which implies taking $\epsilon^0=0$, only the term with the $a$ coefficient contributes to the potential, providing the $\epsilon\,\epsilon^{\prime}$ combination. The other vertex gives the same structure and then we have the $\epsilon\,\epsilon^{\prime}\,\epsilon\,\epsilon^{\prime}$ combination. Hence, the potential can be rewritten as
\begin{equation}
t_{B^*\bar{B}^*}^{\sigma}(\vec{q})=a^2\, \frac{3}{2}\,\left[\frac{1}{f^2}\,\frac{\vec{q}^{\,2}+\frac{m_{\pi}^2}{2}}{1-G(-\vec{q}^{\,2})\,\frac{1}{f^2}(\vec{q}^{\,2}+\frac{m_{\pi}^2}{2})}\right]\,\epsilon_{1\mu}\, \epsilon_{2\nu}^{\prime}\,\epsilon^{\mu}_3\,\epsilon_4^{\prime\,\nu} \, ,
\end{equation}
where we have rewritten the polarization vectors combinations in order to associate the subindices $1\, , 2\, , 3$ and $4$ with $1+2\rightarrow 3+4$. Therefore, the final form of the potential can be written in terms of the spin projectors, Eq. \eqref{eq:projection}, providing
\begin{equation}
t_{B^*\bar{B}^*}^{\sigma}(\vec{q})=a^2\, \frac{3}{2}\,\left[\frac{1}{f^2}\,\frac{\vec{q}^{\,2}+\frac{m_{\pi}^2}{2}}{1-G(-\vec{q}^{\,2})\,\frac{1}{f^2}(\vec{q}^{\,2}+\frac{m_{\pi}^2}{2})}\right]\,(\mathcal{P}^{(0)}+\mathcal{P}^{(1)}+\mathcal{P}^{(2)})\, .
\label{eq:tsigma}
\end{equation}
In Fig. \ref{fig:tsigma} we can see the plot of the $t^{\sigma}_{B^*\bar{B}^*}$ potential, Eq. \eqref{eq:tsigma}, as a function of the transferred momentum $\vec{q}$.


\begin{figure}[htpb]
\centering
\includegraphics[scale=0.5]{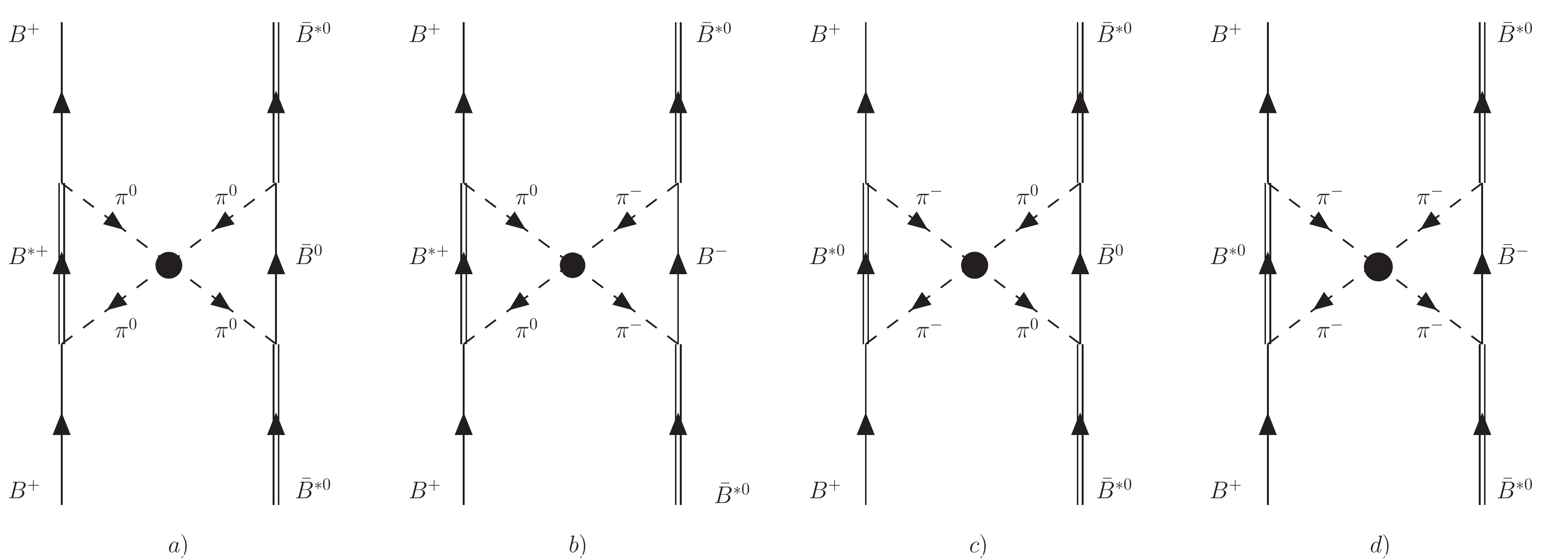}
\caption{Diagrams contributing to the two pion exchange interaction in lowest order for the $B\bar{B}^*\rightarrow B\bar{B}^*$ process in $I=1$.}
\label{fig:diagbbs}
\end{figure}

\begin{figure}[htpb]
\centering
\includegraphics[scale=0.8]{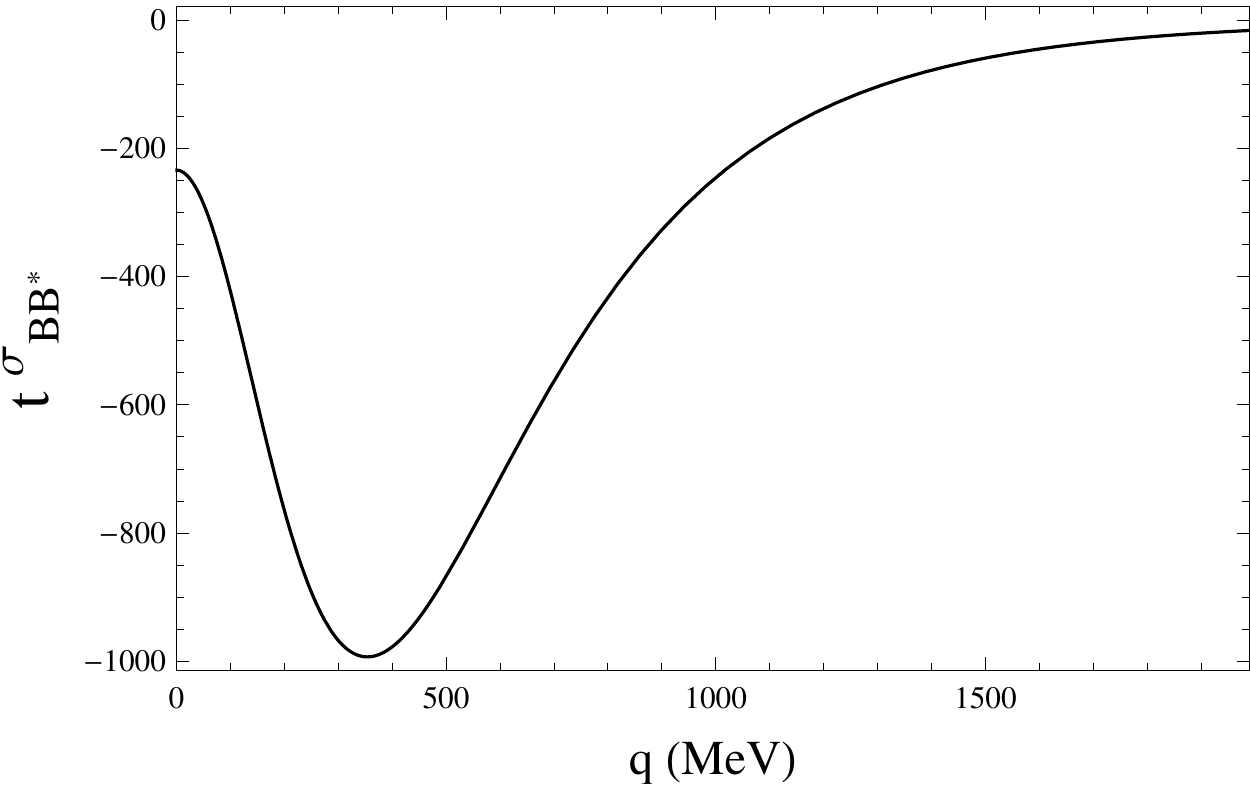}
\caption{Potential $t^{\sigma}_{B\bar{B}^*}$ as a function of the momentum transferred in the process.}
\label{fig:tBbsigma}
\end{figure}

Next, we shall consider the same mechanism, but now for the $B\bar{B}^*$ case. The diagrams for this process are shown in Fig. \ref{fig:diagbbs}. For this case, the potential $t_{B\bar{B}^*}^{\sigma}$ has a difference in comparison with the former case. Now we have two different triangular loops. This implies two $V$ factors in Eq. \eqref{eq:ampl2}, where each factor is associated with each triangular loop. Hence, the potential $t_{B\bar{B}^*}^{\sigma}$ is given by
\begin{equation}
-it_{B\bar{B}^*}^{\sigma}=-i\ V\, \bar{V}\ \frac{3}{2}\ t_{\pi\pi\rightarrow\pi\pi}^{I=0}\ ,
\label{eq:ampl_sigma}
\end{equation}
where $t_{\pi\pi\rightarrow\pi\pi}^{I=0}$ is the isoscalar amplitude defined in Eq. \eqref{eq:isoampl} and $\bar{V}$ is again given by Eq.(29) in \cite{Aceti:2014kja} with trivial changes of masses. The potential $t^{\sigma}_{B\bar{B}^*}$ is plotted in Fig. \ref{fig:tBbsigma}. 

\subsection{The exchange due to the two uncorrelated pions }
\label{uncorrelated}


In this case, the pions are not interacting, then only the diagrams a) and d) of Figs. \ref{fig:diagsigma} and \ref{fig:diagbbs} contribute for the $B^*\bar{B}^*$ and $B\bar{B}^*$ interactions. Details on the evaluation can be found in \cite{Aceti:2014kja}. The amplitude $t_{B^*\bar{B}^*}^{\pi\,\pi}$ can be rewritten in terms of its spin components as

\begin{figure}
  \centering
  \subfigure[]{\label{fig:momentaBsBs}\includegraphics[width=0.3\textwidth]{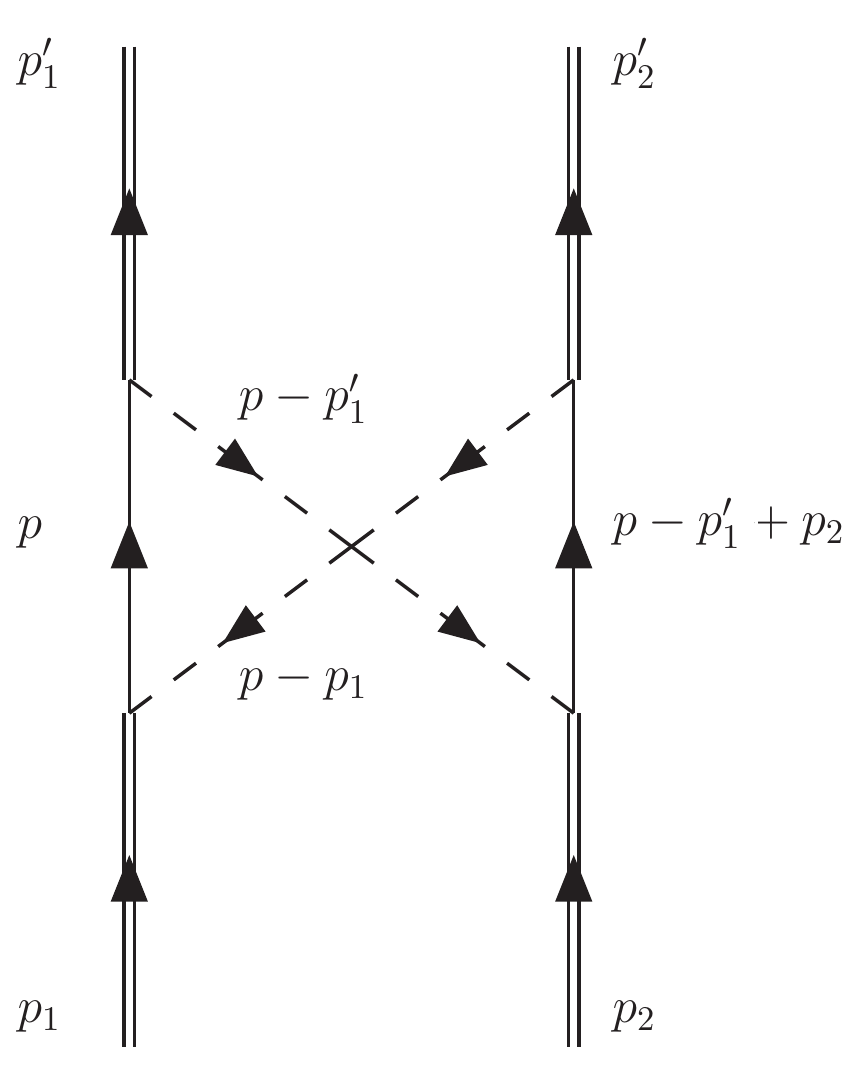}}                
   \subfigure[]{\label{fig:momentaBBs}\includegraphics[width=0.3\textwidth]{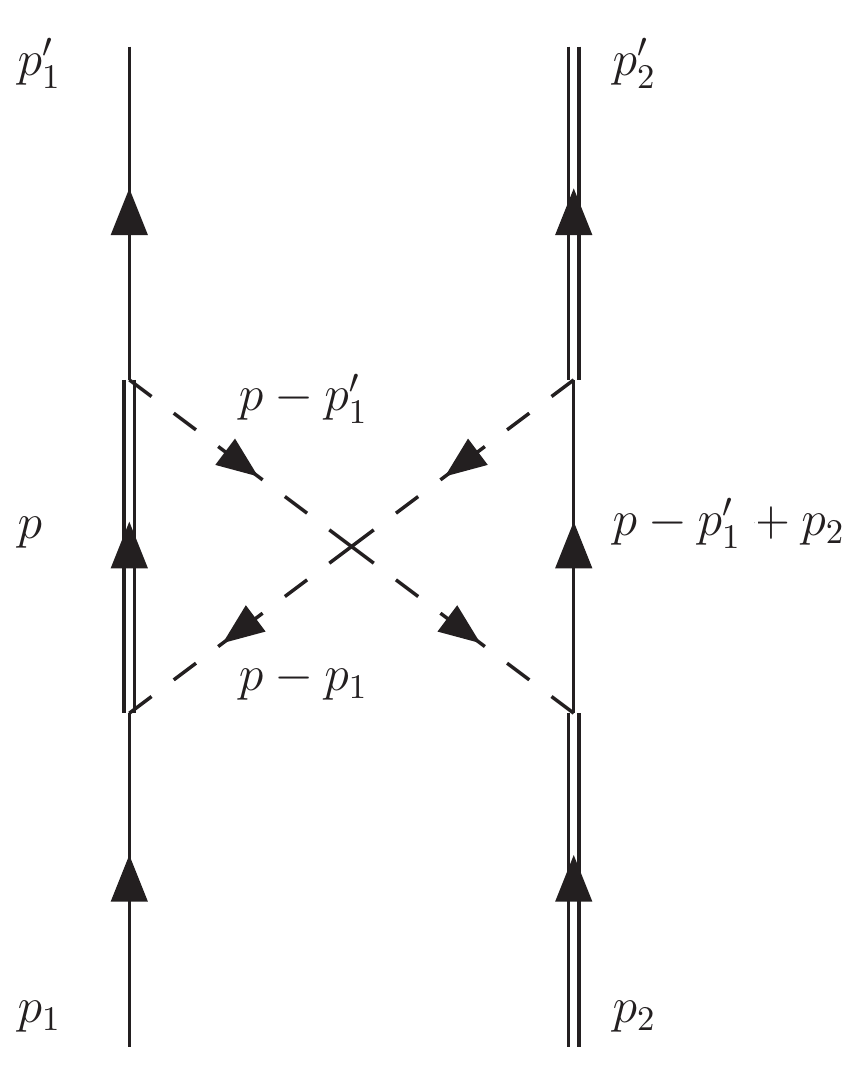}}
  \caption{Momenta assignments in the two uncorrelated pion exchange in $B^*\bar{B}^*\rightarrow B^*\bar{B}^*$ and $B\bar{B}^*\rightarrow B\bar{B}^*$, respectively.}
\label{fig:momenta}
\end{figure}

\begin{equation}
\begin{split}
t_{B^*\bar{B}^*}^{\pi\,\pi}&=\frac{5}{4}\,g_B^4\frac{A}{15}\int\frac{d^3p}{(2\pi)^3}(4\vec{p}^{\,2}-\frac{\vec{q}^{\,2}}{4})^2\, F^2\,\frac{1}{\omega_1+\omega_2}\,\frac{1}{2\omega_1\omega_2}\,\frac{1}{4E_B^2}\,\frac{1}{p_1^0-\omega_1-E_B+i\epsilon}\\&\times\frac{1}{p_1^0-\omega_2-E_B+i\epsilon} \left(1+\frac{E_B+\omega_1+\omega_2-p_1^0}{p_1^0-\omega_1-E_B+i\epsilon}+\frac{E_B+\omega_1+\omega_2-p_1^0}{p_1^0-\omega_2-E_B+i\epsilon}\right)\ ,
\label{eq:ampl_norel4}
\end{split}
\end{equation}
where $A=5$ is associated with spin $J=0$, while $A=2$ is related to the $J=2$ case,  $\omega_1=\sqrt{(\vec{p}+\vec{q}/2)^2+m_{\pi}^2}$, $\omega_2=\sqrt{(\vec{p}-\vec{q}/2)^2+m_{\pi}^2}$ are the energies of the pions and $E_{B}(\vec{p}\,)=\sqrt{\vec{p}^{\ 2}+m_{B}^2}$ is the energy of the $B$ meson. $F(\vec{q\,})$ is a form factor of the type 
\begin{equation}
F=F_1(\vec{p}+\frac{\vec{q}}{2})\, F_2(\vec{p}-\frac{\vec{q}}{2})=\frac{\Lambda^2}{\Lambda^2+(\vec{p}+\frac{\vec{q}}{2})^2}\,\frac{\Lambda^2}{\Lambda^2+(\vec{p}-\frac{\vec{q}}{2})^2}\ ,
\label{eq:ff}
\end{equation}
with $\Lambda=700$ GeV, which is also used later to help the convergence. Note that, according to \cite{xiaoliang}, the coupling $g=M_V/2f_{\pi}$ used in Sec. \ref{vecex} is now replaced by $g_B=(M_{B^*}/M_{K^*})\,g$ to account for the requirements of heavy quark spin symmetry. On the other hand, this correction is automatically implemented in the extrapolation of the vector exchange to the heavy sector (Weinberg-Tomozawa term) because this term is explicitly proportional to the external $B^*$ energies.
\begin{figure}[htpb]
\centering
\includegraphics[scale=0.8]{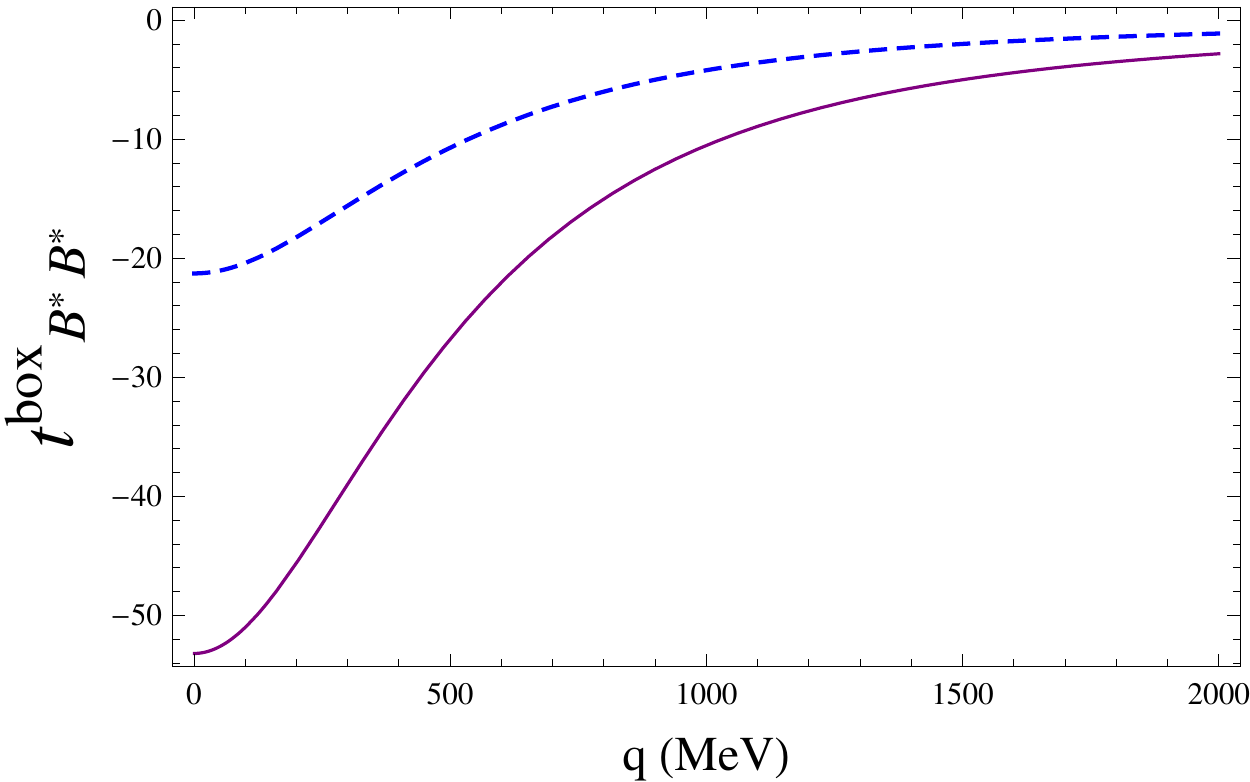}
\caption{Potential $t_{\pi\pi}^{B^*\bar{B}^*}$ for non-interacting pion exchange in the case of $J=0$ (solid line) and $J=2$ (dashed line).}
\label{fig:pions}
\end{figure}

In Fig. \ref{fig:pions} we can see the amplitude for the two spin cases as a function of the momentum transfer.

\begin{figure}[htpb]
\centering
\includegraphics[scale=0.65]{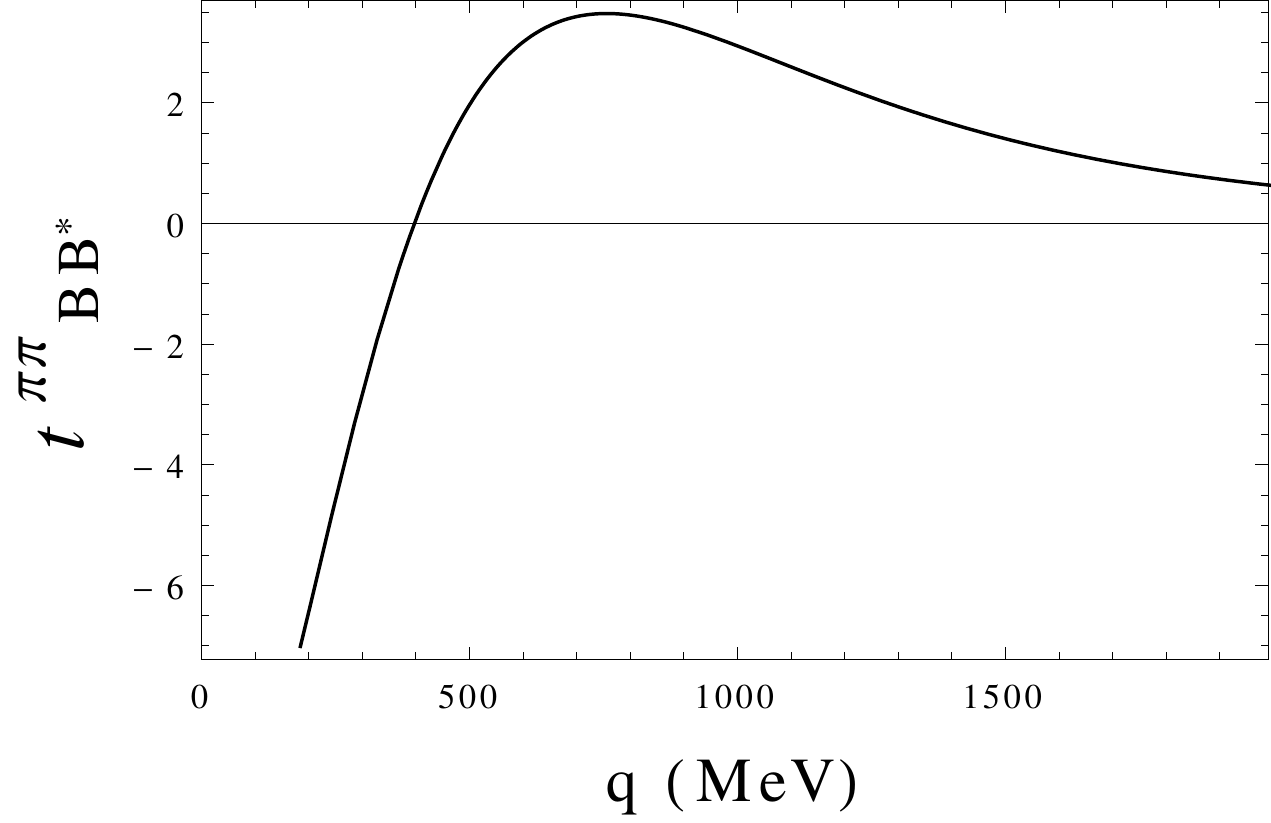}
\caption{Potential $t_{B\bar{B}^*}^{\pi\,\pi}$ for non-interacting pion exchange as a function of the momentum transferred in the process.}
\label{fig:pionex}
\end{figure}

For the $B\bar{B}^*$ case we find
\begin{equation}
\begin{split}
t_{B\bar{B}^*}^{\pi\,\pi}&=-\frac{5}{4}g_B^4\,\frac{1}{2}\,\vec{\epsilon}\ '\ \vec{\epsilon}\ ''\int\frac{d^3p}{(2\pi)^3}\, (\vec{p}^{\ 2}-\vec{q}^{\ 2})\left[(4\vec{p}^{\ 2}-\frac{\vec{q}^{\ 2}}{4})-\frac{1}{\vec{q}^{\ 2}}\left[(2\vec{p}\,\vec{q}\,)^2-\frac{\vec{q}^{\ 4}}{4}\right]\right]\,\frac{F^2}{\omega_1+\omega_2}\,\frac{1}{2\omega_1\omega_2}\\&\times\frac{1}{2E_B}\,\frac{1}{2E_V}[\omega_1^2+\omega_2^2+\omega_1\omega_2-(\omega_1+\omega_2)(2p_1^0-E_{B^*}-E_B)+(p_1^0-E_{B^*})(p_1^0-E_B)]\\&\times\frac{1}{p_1^0-\omega_1-E_{B^*}+i\epsilon}\,\frac{1}{p_1^0-\omega_1-E_B+i\epsilon}\,\frac{1}{p_1^0-\omega_2-E_{B^*}+i\epsilon}\,\frac{1}{p_1^0-\omega_2-E_B+i\epsilon}\ ,
\label{eq:ampl_norel2}
\end{split}
\end{equation}
where $E_{B^*}(\vec{p}\,)=\sqrt{\vec{p}^{\ 2}+m_{B^*}^2}$ is the energy of the $B^*$ meson. The amplitude $t_{B\bar{B}^*}^{\pi\,\pi}$ as a function of the momentum transfer is plotted in Fig. \ref{fig:pionex} 

\section{Iterated exchange of two light mesons}

\begin{figure}[htpb]
\centering
\includegraphics[scale=0.5]{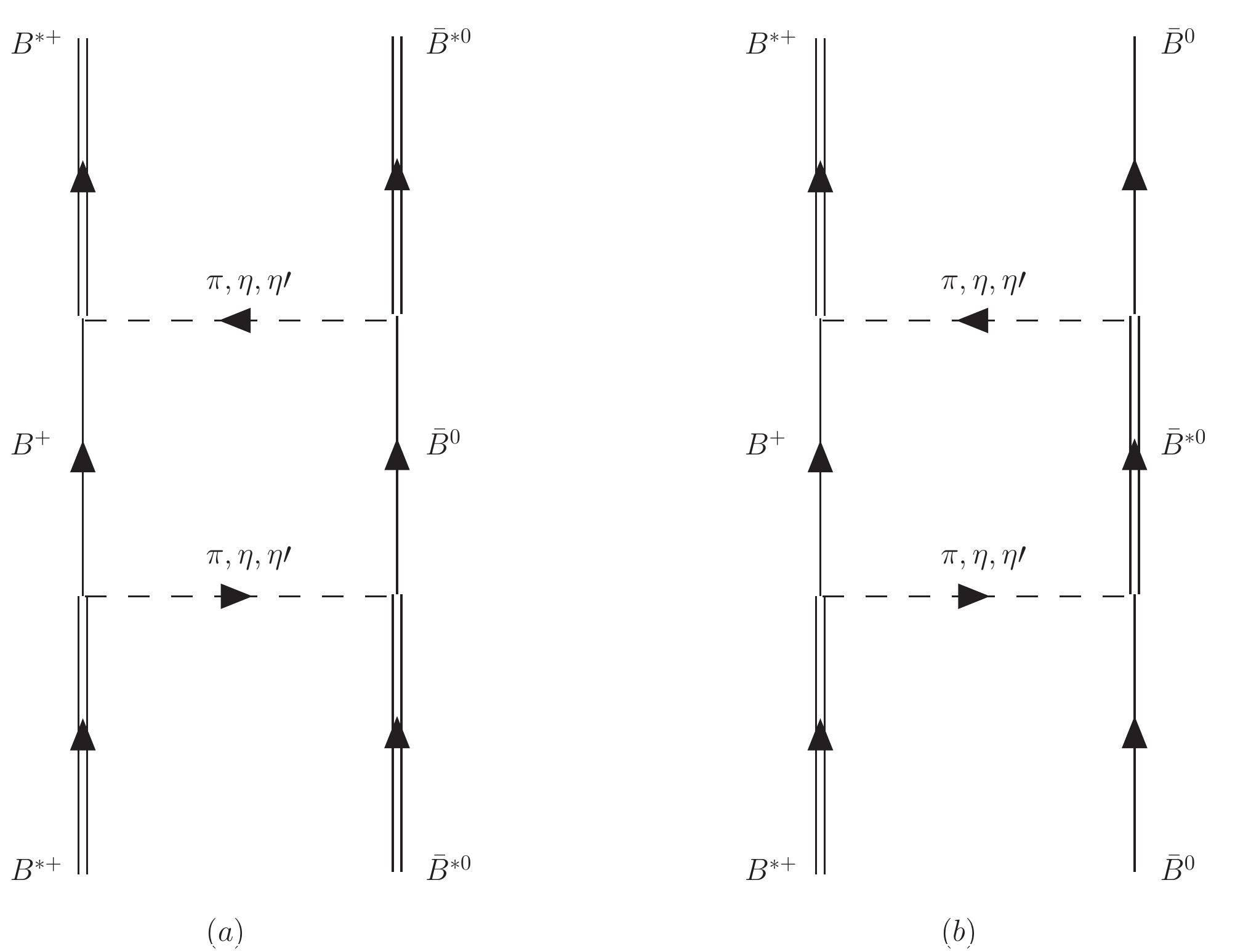}
\caption{Iterated exchange of two light mesons for the $B^*\bar{B}^*$ (a) and $B\bar{B}^*$ (b) cases.}
\label{fig:boxdiagram}
\end{figure}

In this section we evaluate the contribution coming from the iterated exchange of two light mesons, shown in Fig. \ref{fig:boxdiagram} in the case of the $B^*\bar{B}^*$ (a) and of the $B\bar{B}^*$ (b) interactions. 

In the case of $B^*\bar{B}^*$, the details of the calculation can be found in Sec. C of Ref. \cite{Aceti:2014kja} and they lead to the following expression for the amplitude:
\begin{equation}
t^{box}_{B^*\bar{B}^*}=\frac{1}{4}t^{box}_{\pi\pi}+\frac{1}{9}t^{box}_{\eta\eta}+\frac{1}{36}t^{box}_{\eta'\eta'}-\frac{1}{3}t^{box}_{\pi\eta}-\frac{1}{6}t^{box}_{\pi\eta'}+\frac{1}{9}t^{box}_{\eta\eta'}\ ,
\label{eq:boxBsBs1}
\end{equation}
where 
\begin{eqnarray}
t^{box}_{ij}&=&g_B^{4}S_{J}\int \frac{d^{3}p}{(2\pi)^{3}}\,\vec{p}^{\ 4}\,F^2\,\frac{1}{m_{D^*}+\omega_{1}-E_{B}(\vec{p})\pm i\epsilon}\,\frac{1}{m_{B^*}+\omega_{2}-E_{B}(\vec{p})\pm i\epsilon}\nonumber \\
&\times&\frac{1}{(E_{B}(\vec{p}\,))^2}\Big (\frac{1}{2\omega_1\omega_2}\,\frac{1}{\omega_1+\omega_2}\,\frac{\textrm{Num}}{m_{B^{*}}-\omega_1-E_{B}(\vec{p}\,)+i\epsilon}\,\frac{1}{m_{B^{*}}-\omega_2-E_{B}(\vec{p}\,)+i\epsilon} \nonumber \\
&+&\frac{1}{E_B(\vec{p}\,)-m_{B^*}+\omega_1+i\epsilon}\,\frac{1}{E_B(\vec{p}\,)-m_{B^*}+\omega_2+i\epsilon}\,\frac{1}{2m_{B^*}-2E_{B}(\vec{p}\,)+i\epsilon}\Big )\ ,
\label{eq:boxBsBs2}
\end{eqnarray}
where $i\,j=\pi,\, \eta,\, eta^{\prime}$ and $\omega_1$ and $\omega_2$ are their energies of the two light mesons exchanged, 
\begin{equation}
S_J=\begin{cases} 
\frac{4}{3}\ \ \ \ \ \ J=0\\ 
\\
\frac{8}{15}\ \ \ \ \ \ J=2\ ,
\end{cases}
\end{equation}
and 
\begin{equation}
\textrm{Num}=-(\omega_1^2+\omega_2^2+\omega_1\omega_2)+(m_{B^*}-E_{B}(\vec{p}\,))^2\ .
\end{equation}
The former calculation has been done at threshold. The momentum transfer dependence on $\vec{q}$ can be obtained easily from Eq. \eqref{eq:boxBsBs2} by taking for the initial and final states four-momenta $p_1=(p_1^0,\vec{q}/2)$, $p_2=(p_2^0,-\vec{q}/2)$, $p_3=(p_3^0,-\vec{q}/2)$ and $p_4=(p_4^0,\vec{q}/2)$ ($p_3,\, p_4$ momenta of the two final $B^*$).

\begin{figure}[htpb]
\centering
\includegraphics[scale=0.8]{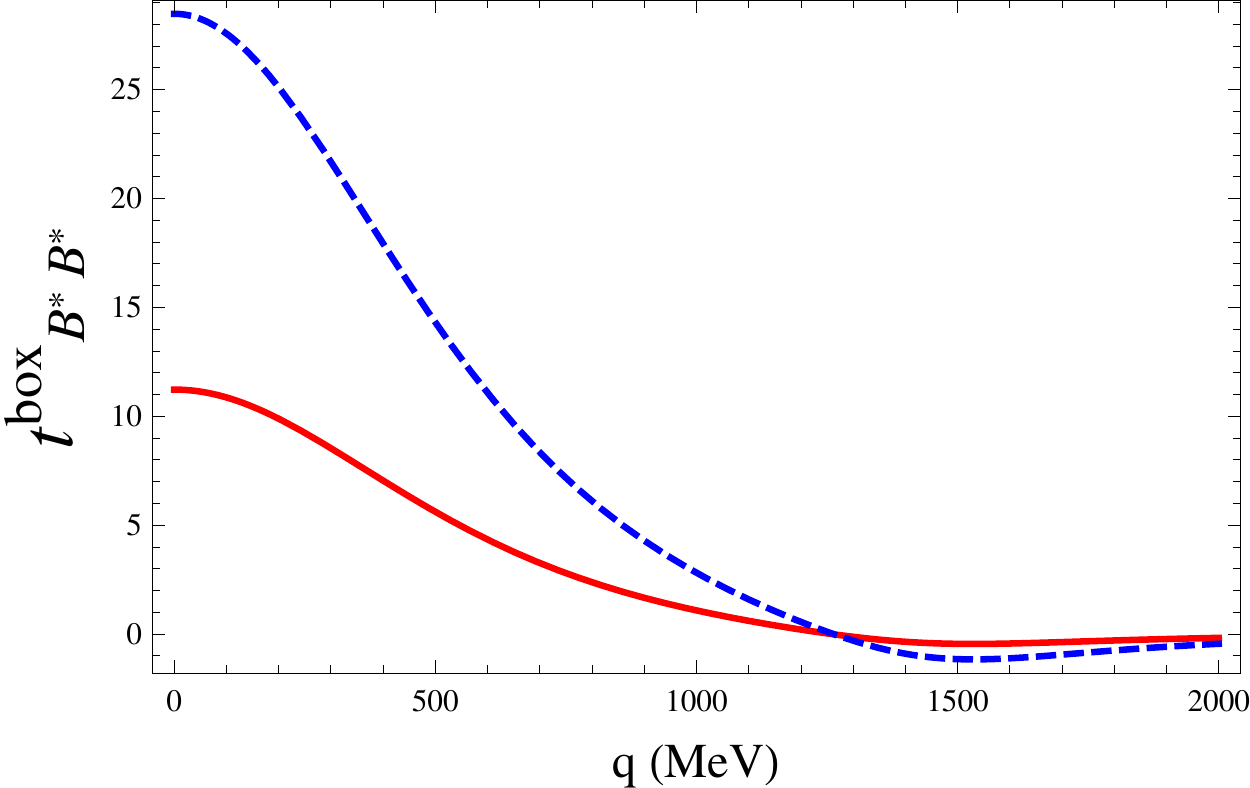}
\caption{Amplitude $t^{box}_{B^*\bar{B}^*}$ as a function of the momentum transferred in the process for the case $J=0$ (dashed line) and $J=2$ (solid line).}
\label{fig:boxbsbs}
\end{figure}

In Fig. \ref{fig:boxbsbs} the amplitude $t^{box}_{B^*\bar{B}^*}$ is plotted as a function of the momentum transferred $\vec{q}$ for the case $J=0$ (dashed line) and $J=2$ (solid line). 

With a similar procedure we can obtain the amplitude in the case of $B\bar{B^*}$, using again the Lagrangian of Eq. \eqref{eq:VPPlag}. We find
\begin{equation}
t^{box}_{B\bar{B}^*}=\frac{1}{4}\tilde{t}^{box}_{\pi\pi}+\frac{1}{9}\tilde{t}^{box}_{\eta\eta}+\frac{1}{36}\tilde{t}^{box}_{\eta'\eta'}-\frac{1}{3}\tilde{t}^{box}_{\pi\eta}-\frac{1}{6}\tilde{t}^{box}_{\pi\eta'}+\frac{1}{9}\tilde{t}^{box}_{\eta\eta'}\ ,
\label{eq:boxBBs1}
\end{equation}
where
\begin{eqnarray}
\label{eq:boxBBs2}
\tilde{t\,}^{box}_{ij}&=&g_B^{4}\frac{1}{3}\,\vec{\epsilon}\cdot\vec{\epsilon}^{\prime}\int\frac{d^3p}{(2\pi)^3}\,\vec{p\,}^4\,F^2\,\frac{1}{E_{B^*}(\vec{p}\,)}\,\frac{1}{E_{B}(\vec{p}\,)}\,\frac{1}{E_{B^*}(\vec{p}\,)+\omega_1-E_{B}(\vec{p}\,)\pm i\epsilon}\nonumber\\
&\times&\frac{1}{E_{B^*}(\vec{p}\,)+\omega_1-E_{B}(\vec{p}\,)\pm i\epsilon}\frac{1}{E_{B^*}(\vec{p}\,)+\omega_2-E_{B}(\vec{p}\,)\pm i\epsilon}\Big (\frac{1}{2\omega_1\omega_2}\nonumber\\
&\times&\frac{1}{\omega_1+\omega_2}\,\frac{1}{E_B(\vec{p}\,)-\omega_1-E_{B^*}(\vec{p}\,)+i\epsilon}\,\frac{\textrm{Num}^{\prime}}{E_B(\vec{p}\,)-\omega_2-E_{B^*}(\vec{p}\,)+i\epsilon}\nonumber\\
&+&\frac{1}{E_B(\vec{p}\,)+\omega_1-E_{B^*}(\vec{p}\,)-i\epsilon}\,\frac{1}{E_B(\vec{p}\,)+\omega_2-E_{B^*}(\vec{p}\,)-i\epsilon}\nonumber\\
&\times&\frac{1}{M_{B}-E_B(\vec{p}\,)+M_{B^*}-E_{B^*}(\vec{p}\,)+i\epsilon}\,\Big )\ ,
\end{eqnarray}
with $i,j=\pi,\eta,\eta^{\prime}$. The numerator $\textrm{Num}^{\prime}$ in Eq. \eqref{eq:boxBBs2} is given by
\begin{equation}
\begin{split}
\textrm{Num}^{\prime}&=-(\omega_1^2+\omega_2^2+\omega_1\omega_2)+(\omega_1+\omega_2)(M_B+E_B-M_{B^*}-E_{B^*})\\
&\times(M_B-E_{B^*})(M_{B^*}-E_{B})\ .
\end{split}
\end{equation}
$E_{B}$, $E_{B^*}$, $\omega_1$ and $\omega_2$ are already defined in Section \ref{uncorrelated}.

\begin{figure}[htpb]
\centering
\includegraphics[scale=0.8]{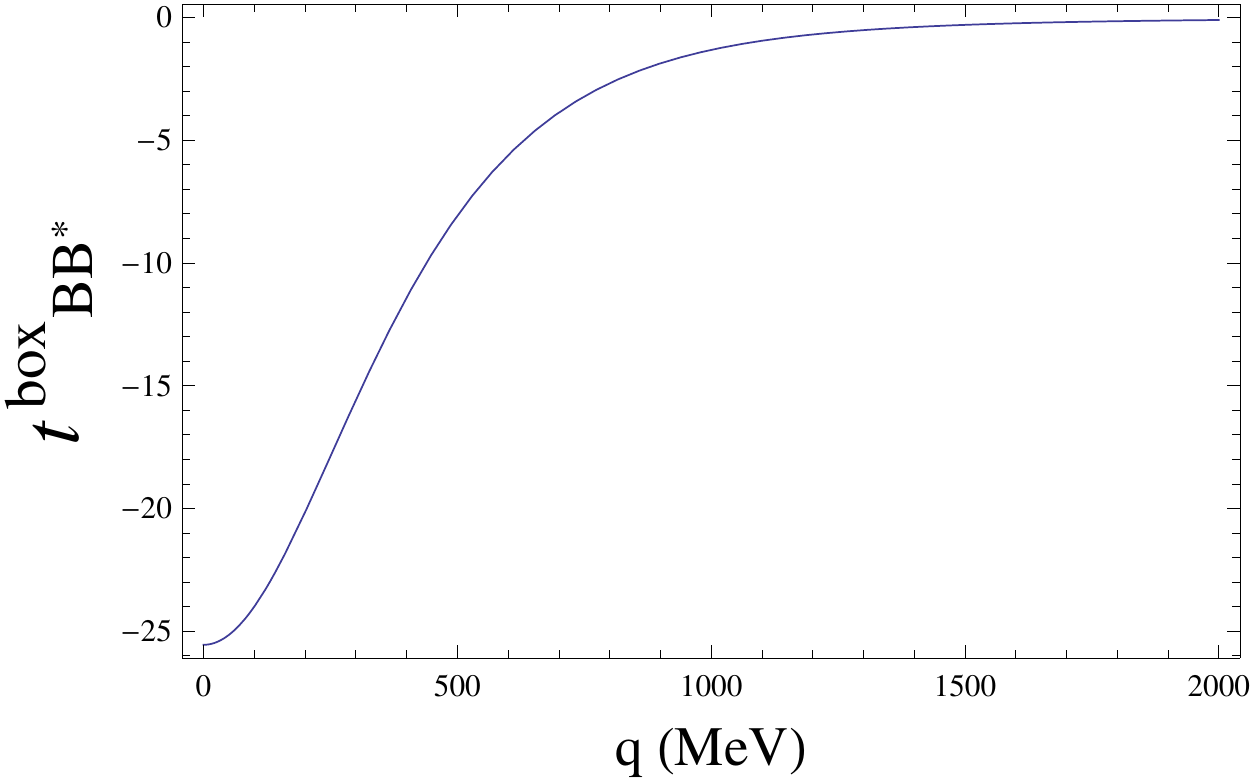}
\caption{Amplitude $t^{box}_{B\bar{B}^*}$ as a function of the momentum transferred in the process.}
\label{fig:boxbbs}
\end{figure}

The potential $t_{B\bar{B}^*}^{box}$ is plotted in Fig. \ref{fig:boxbbs} as a function of the tranferred momentum $\vec{q}$.

\section{Results}

\begin{figure}[htpb]
\centering
\includegraphics[scale=0.8]{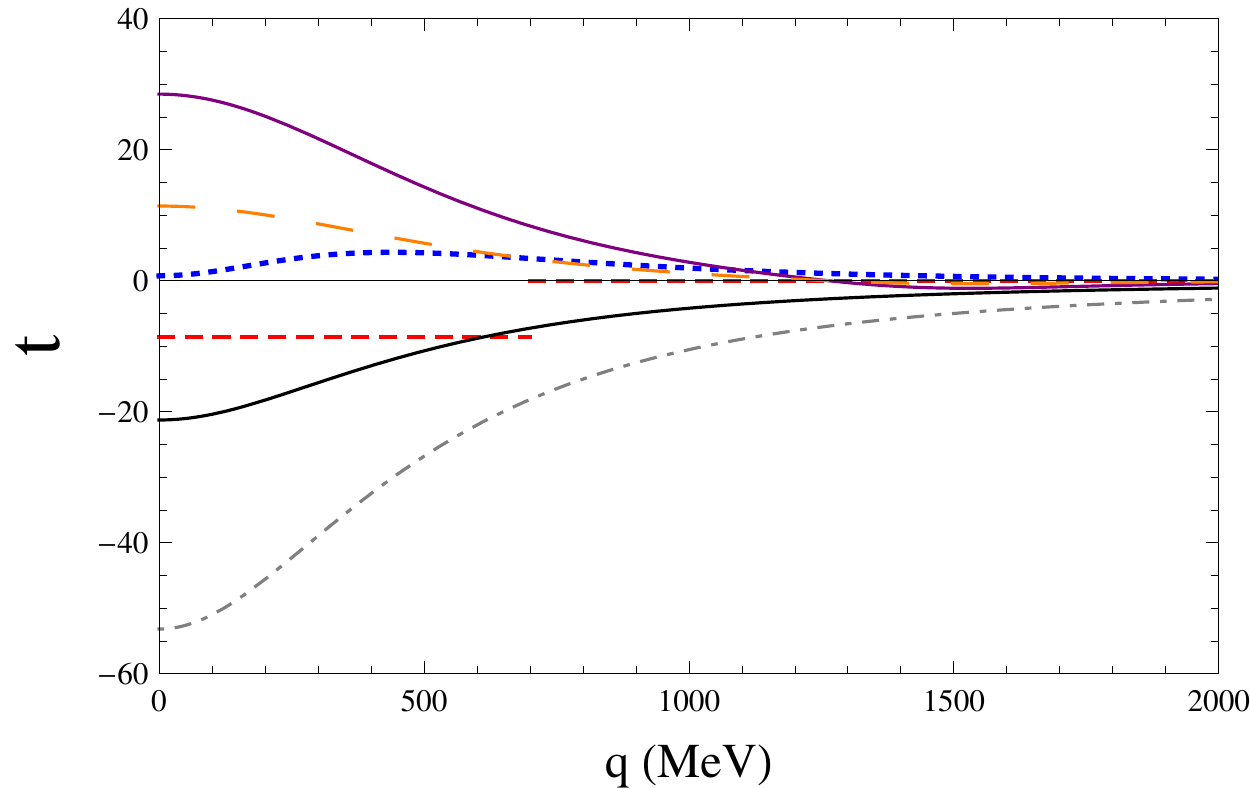}
\caption{Comparison between the potentials  $t_{B^*\bar{B}^*\to B^*\bar{B}^*}$ (small dashed line, vector exchange Eq. \eqref{eq:tbb2}), $t_{B^*\bar{B}^*}^{\sigma}$ (dotted line, Eq. \eqref{eq:tsigma}), $t_{B^*\bar{B}^*}^{\pi\pi}$ for $J=0$ (dotted dashed line, Eq. \eqref{eq:ampl_norel4}) and $J=2$ (solid line), $t^{box}_{B^*\bar{B}^*}$ for $J=0$ (solid thick line, Eq. \eqref{eq:boxBsBs1}) and $J=2$ (large dashed line) as functions of the momentum transferred in the process.}
\label{fig:tcompare}
\end{figure}

\begin{figure}[htpb]
\centering
\includegraphics[scale=0.8]{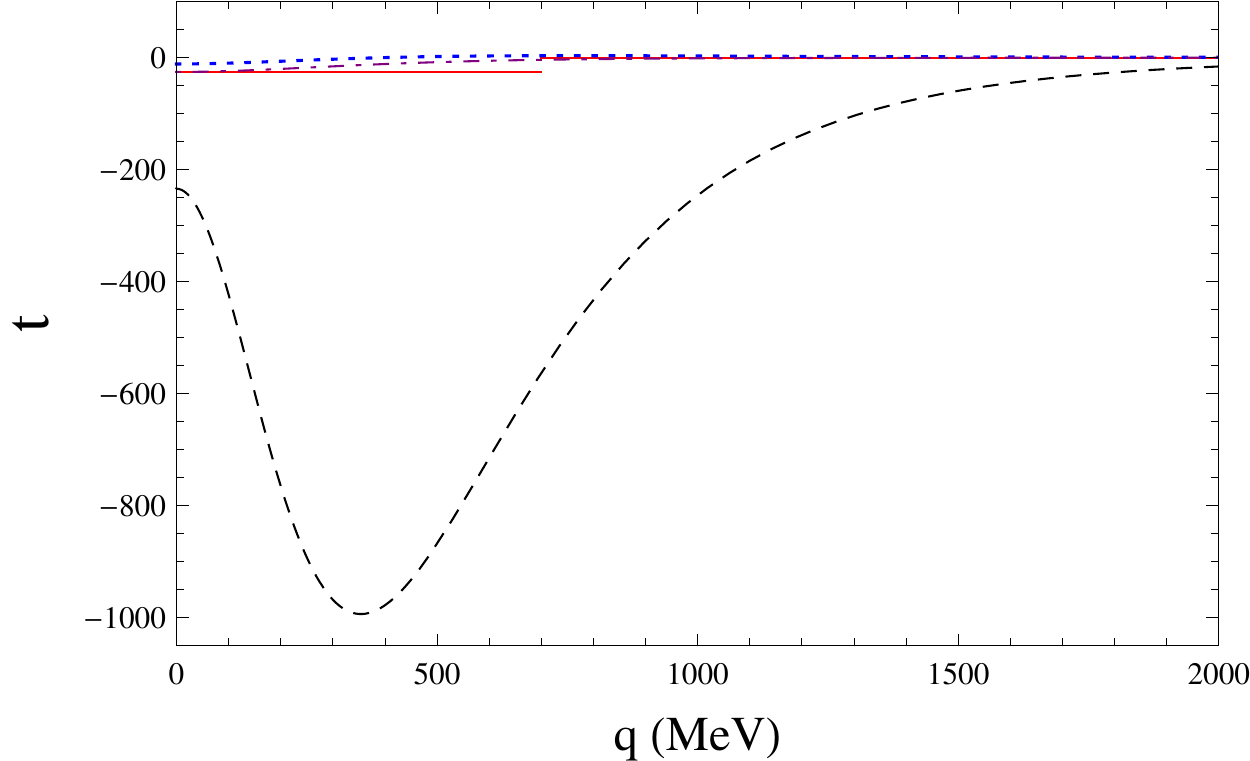}
\caption{Comparison between the potentials  $t_{B\bar{B}^*\to B\bar{B}^*}$ (solid line, Eq. \eqref{eq:VVPPamp}), $t_{B\bar{B}^*}^{\sigma}$ (dashed line, Eq. \eqref{eq:ampl_sigma}), $t_{B\bar{B}^*}^{\pi\pi}$(dotted line, Eq. \eqref{eq:ampl_norel2}), $t^{box}_{B\bar{B}^*}$ (dotted dashed line, Eq. \eqref{eq:boxBBs1}) as functions of the momentum transferred in the process.}
\label{fig:tcompare2}
\end{figure}

After we have calculated the amplitudes of all the processes contributing to the $B^*\bar{B}^*$ and $B\bar{B}^*$ interactions, we want to make a rough estimate of the strength of each potential. This is done by evaluating the integral $\int{d^3q\,V(q)}$ in order to take into account the contributions coming from the exchange of light mesons and use them to obtain an effective potential $V_{eff}$. We will follow a simple strategy to account for the different potentials. We will get the strength $\int{d^3q\,V_i\,(q)}$ for all the potentials exchanging light mesons and sum them. Then we convert the sum into an effective potential of the type of the vector exchange, 

\begin{equation}
V_{eff}\theta (q_{max}-|\vec{q\,}|)\theta(q_{max}-|\vec{q\,}^{\prime}|)\, ,
\end{equation}
where $q_{max}$ is the maximum momentum used in the loops in Eq. \eqref{eq:1striemann} (see Eq. \eqref{eq:potheta}), such that $\int_{q<q_{max}}d^3q\,V_{eff}$ is equal to the sum of $\int{d^3q\,V_i\,(q)}$.

Then, we take as potential in our case this effective potential plus the one coming from vector exchange. Both are of the type of Eq. \eqref{eq:potheta} and can then be used in the Bethe-Salpeter equation with the same $G$ function (Eq. \eqref{eq:1striemann}), regularized with the cut off $q_{max}$.

On the other hand, the value of the strength depends on the value of the upper limit of the integral $\int{d^3q\,V(q)}$. For this reason we calculated the effective potential $V_{eff}$ using values of this limit for the light meson exchange potential varying from $700$ to $1100$ MeV for both $B^*\bar{B}^*$ and $B\bar{B}^*$ interactions. Changing the upper limit in $\int{d^3q\,V_i\,(q)}$ introduces large uncertainties in the approach concerning the final potential. The strength of the final potential, summing $V_{eff}$ and the vector exchange, can be a factor $2.4-14.5$ times the one of the vector exchange alone for the case of $B^*\bar{B}^*$ with $J=0$, while for $J=2$ we find a factor $1.2-5.2$. For the case of $B\bar{B}^*$ the factor varies between $30$ and $64$.

In the following we study the shape of $|T_{11}|^2$ for both $B^*\bar{B}^*$ and $B\bar{B}^*$ cases. As we will discuss in detail, both amplitudes show a clear peak and the large uncertainties on the potential do not affect drastically its position, which justifies a posteriori the approach followed indulging in large uncertainties.


\subsection{$B\bar{B}^*$ case}

\begin{figure}[htpb]
\centering
\includegraphics[scale=0.65]{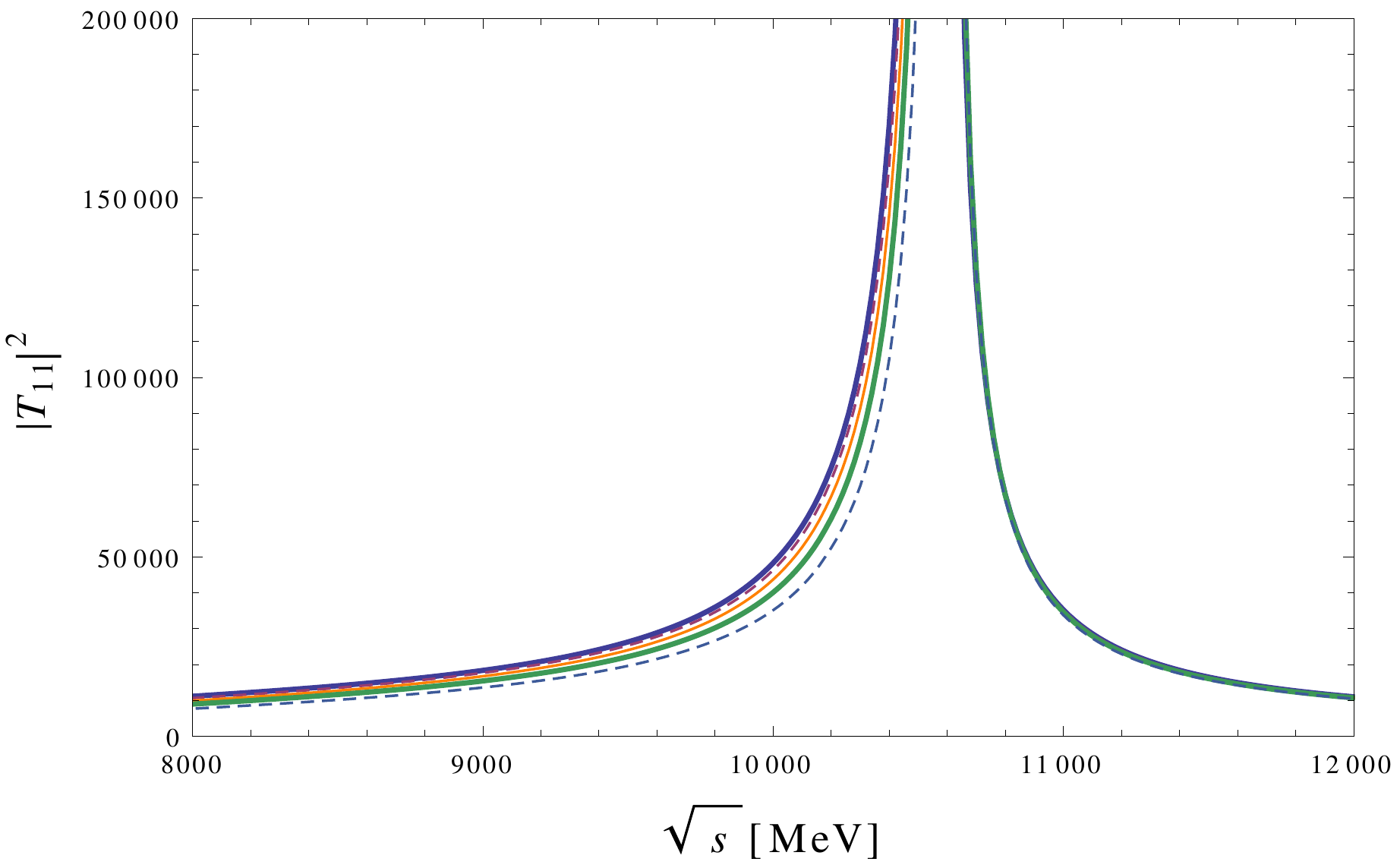}
\caption{$|T_{11}|^{2}$ as a function of the $\sqrt{s}$ center of mass energy for the case of $B\bar{B}^*$. Each curve is associated with a value of the integration limit: $700$ MeV, $800$ MeV, $900$ MeV, $1000$ MeV, $1100$ MeV. The peak moves from right to left as the integration limit increases.}
\label{fig:peakBBs}
\end{figure}

In this case, we are interested in studying the $T$ matrix for the channels: $B\bar{B}^*$, $\eta_b\,\rho$ and $\pi\,\Upsilon$. We evaluated the transition matrix $T$ between those channels for values of $\sqrt{s}$ around 10600 MeV. In order to do this, we use the dimensional regularization formula for the loop function $G$, given by Eq. \eqref{eq:loopexdm}. To obtain reasonable values of the subtraction constants in each channel we proceed as follows: we take a cut off $q_{max}$, then we find a subtraction constant that provides at threshold the same $G$ function obtained with the cut off method. Here we are taking $q_{max}=700$ MeV, for which we find $\alpha_{B\bar{B}^*}=-2.79$, $\alpha_{\eta_b\rho}=-3.56$ and $\alpha_{\pi\Upsilon}=-3.78$.

In Ref. \cite{Aceti:2014kja,Aceti:2014uea} the changes in the position of the peak of the $T$ matrix due to the variation of $q_{max}$ were studied. However, in the current case, the changes due to this parameter are smaller than the ones due to the variations of the upper limit of the integral $\int{d^3q\,V(q)}$ used to estimate $V_{eff}$. In Fig. \ref{fig:peakBBs} the shape of $|T_{11}|^2$, the component of the $T$ matrix that describes the transition $B\bar{B}^*\rightarrow B\bar{B}^*$, for different values of the integration limit, is depicted. As can be seen, even choosing values of the limit between $700$ and $1100$ MeV, the effect on the binding and the width is small. As a result, we find that the position of the peak moves slightly to higher energies for decreasing values of the upper limit and it is seen in the range of $10587-10601$ MeV. This values are very close to what was found by the Belle collaboration, $M_{Z_b(10610)}=(10608.4\pm2.0)$ MeV. It is worth noting that both the $\eta_b\,\rho$ and $\pi\,\Upsilon$ channels are open for decays, and this gives a width between $1.6$ and $3$ MeV, with bigger widths corresponding to lower values of the integration limit. The experimental value reported by Belle collaboration is $\Gamma_{Z_b(10610)}=(15.6\pm 2.5)$ MeV. 


\subsection{$B^*\bar{B}^*$ case}


\begin{figure}[htpb]
\centering
\includegraphics[scale=0.65]{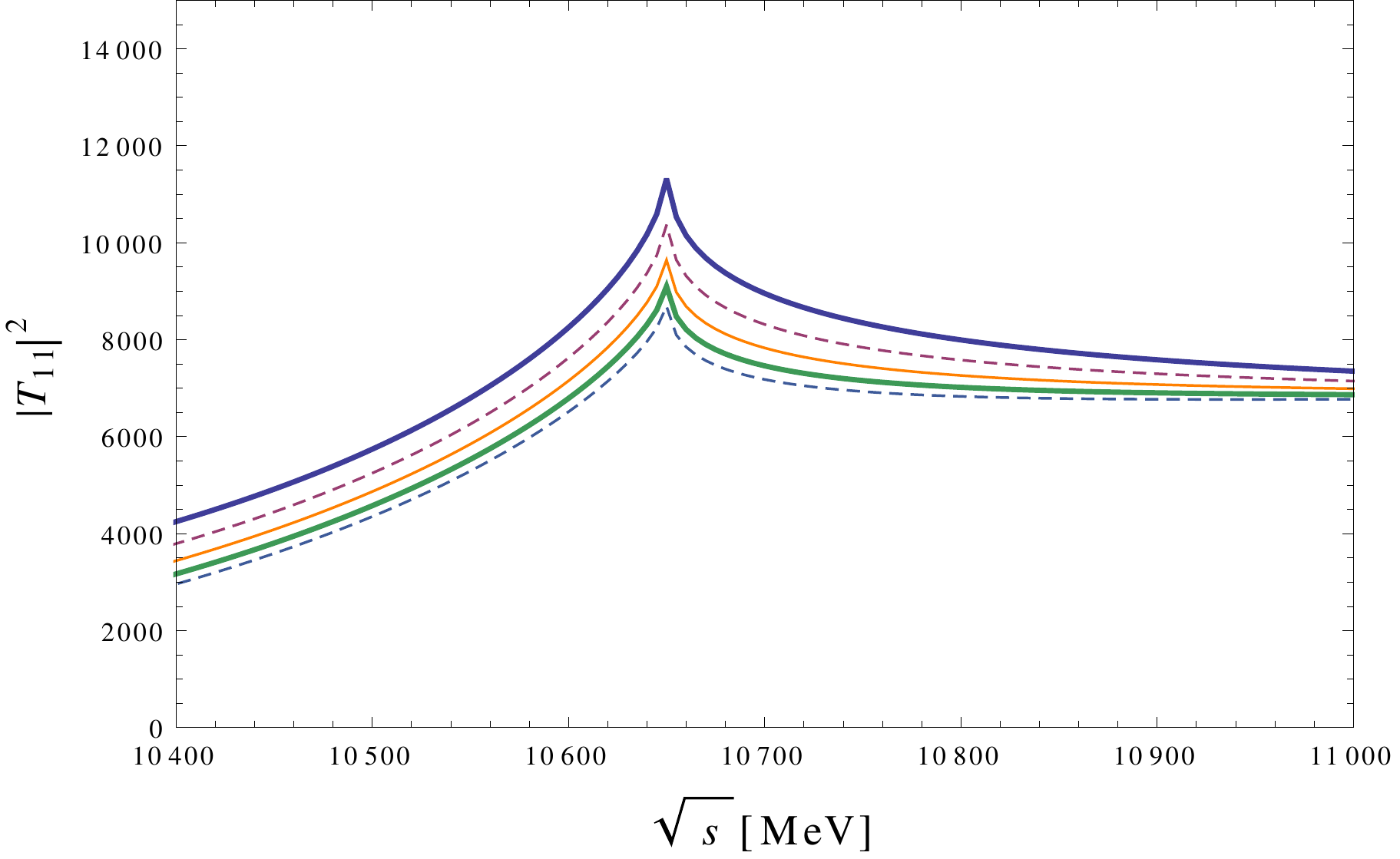}
\caption{$|T_{11}|^{2}$ as a function of the $\sqrt{s}$ center of mass energy for the case of $B^*\bar{B}^*$ for $J=0$. Each curve is associated with a value of the integration limit: $700$ MeV, $800$ MeV, $900$ MeV, $1000$ MeV, $1100$ MeV. The peak moves from bottom to top as the integration limit increases.}
\label{fig:peakBsBs0}
\end{figure}

\begin{figure}[htpb]
\centering
\includegraphics[scale=0.65]{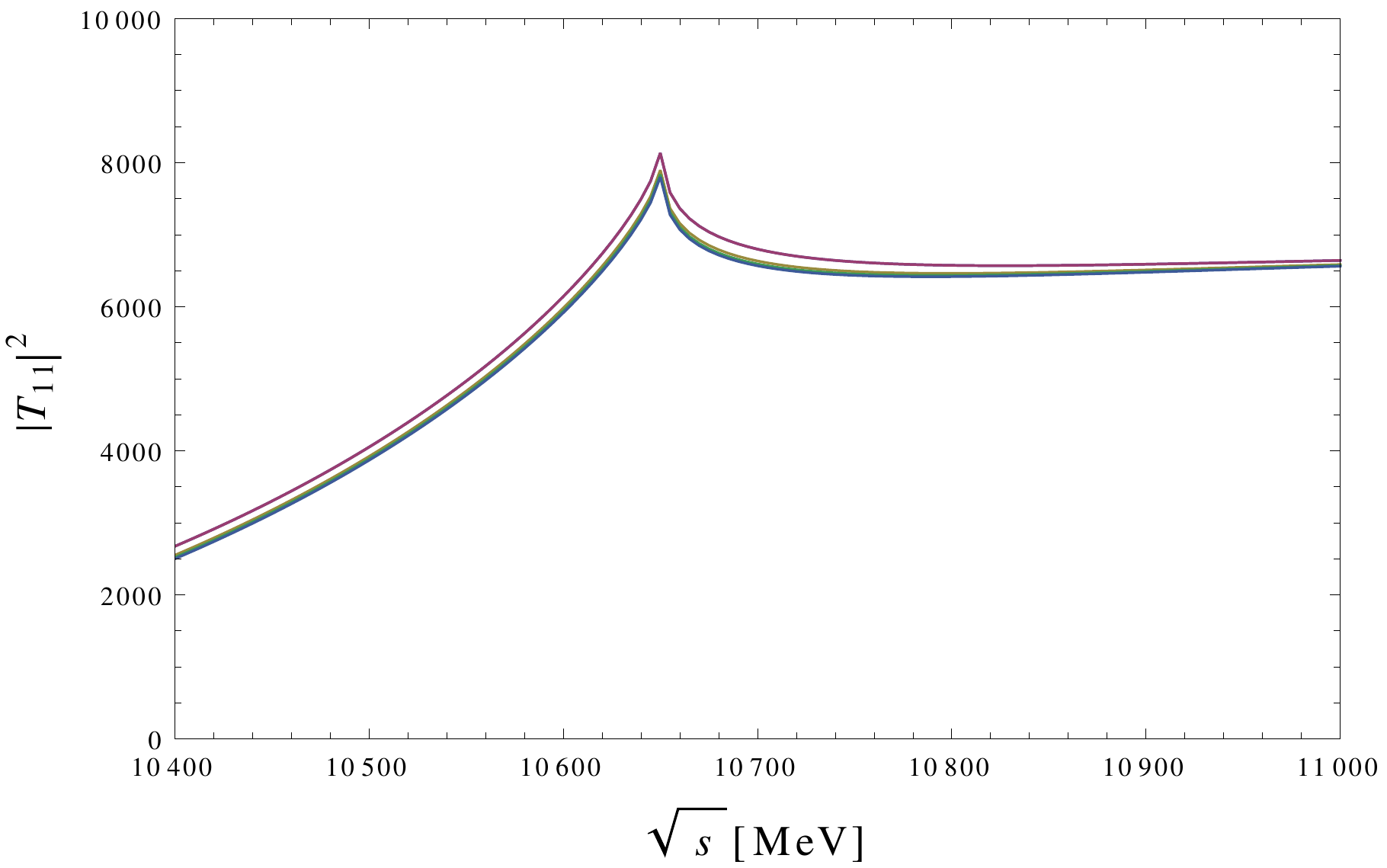}
\caption{$|T_{11}|^{2}$ as a function of the $\sqrt{s}$ center of mass energy for the case of $B^*\bar{B}^*$ for $J=2$. Each curve is associated with a value of the integration limit: $700$ MeV, $800$ MeV, $900$ MeV, $1000$ MeV, $1100$ MeV. The peak moves sligthtly from bottom to top as the integration limit increases.}
\label{fig:peakBsBs2}
\end{figure}

For this case we have two channels: $B^*\bar{B}^*$ and $\rho\, \Upsilon$. Again, we use the dimensional regularization form of the loop function $G$, Eq. \eqref{eq:loopexdm}, with $\mu=1500$ MeV and the subtraction constants $\alpha_{B^*\bar{B}^*}=-2.79$ and $\alpha_{\rho \Upsilon}=-3.56$, corresponding to a cut off value equal to $q_{max}=700$ MeV.

Fig. \ref{fig:peakBsBs0} shows the shape of $|T_{11}|^2$, which means the component of the $T$ matrix that describes the transition from $B^*\bar{B}^*$ to itself, for different values of the integration limit plotted as a function of the center of mass energy, $\sqrt{s}$, of the system. This peak corresponds to spin $J=0$. In Fig \ref{fig:peakBsBs2}, we show the shape of $|T_{11}|^2$ for the $J=2$ case, again for different values of the integration limit. It is important to emphasize that, according to Eq. \eqref{eq:trhoupsilon}, there is no contribution in the transition matrix $T$ from $B^*\bar{B}^*$ to $\rho\,\Upsilon$ channel for spin $J=1$. In this case, $B^*\bar{B}^*$ stands as a single channel.

From these figures we can see that the variations of the integration limit cause no effect to the peak position, as we already noted in the $B\bar{B}^*$ case. It is interesting to note that, even with the large uncertainties in the potential admitted, we always find a structure for the peak of $|T_{11}|^2$ which corresponds clearly to a cusp. Whether to call this a resonant state or not it is a question of criterion. We should however note that the $a_0(980)$ appears in the experiments (or in the theories) \cite{Rubin:2004cq,Oller:1998hw} as a cusp and is universally accepted as a resonance. Our findings, obtained a cusp for the $|T_{11}|^2$ amplitude in this case, would come to support the claims of the former works \cite{Bugg:2011jr,Swanson:2014tra}.

For the sake of completeness, we repeat the calculation considering the spin $J=1$ case. Here we have a single channel problem, 

\begin{equation}
T_{11}= \frac{\tilde{t}_{B^{*}\bar{B}^{*}\rightarrow B^{*}\bar{B}^{*}}}{1-\tilde{t}_{B^{*}\bar{B}^{*}\rightarrow B^{*}\bar{B}^{*}}\,G_{B^*\bar{B}^*}}\, ,
\end{equation} 
where $G_{B^*\bar{B}^*}$ is the loop function defined by Eq. \eqref{eq:loopexdm} for the $B^*\bar{B}^*$ channel, while $\tilde{t}_{B^{*}\bar{B}^{*}\rightarrow B^{*}\bar{B}^{*}}$ is the $B^{*}\bar{B}^{*}\rightarrow B^{*}\bar{B}^{*}$ vector exchange potential already defined in Eq. \eqref{eq:tbb2}, plus the contribution from $V_{eff}$ due to the exchange of two interacting pion exchange. In this case, we saw that the noninteracting pion exchange vanished, and the interacting two pion exchange was also small (see Fig. \ref{fig:tsigma}), smaller than the vector exchange (see Fig. \ref{fig:tcompare}), in all range. This is why, in this case, in order to play with uncertainties we follow the strategy of Refs. \cite{Aceti:2014kja,Aceti:2014uea} and we change the range of the vector exchange potential, by changing the cut off $q_{max}$ to values from $700$ to $1100$ MeV.
\begin{figure}[htpb]
\centering
\includegraphics[scale=0.7]{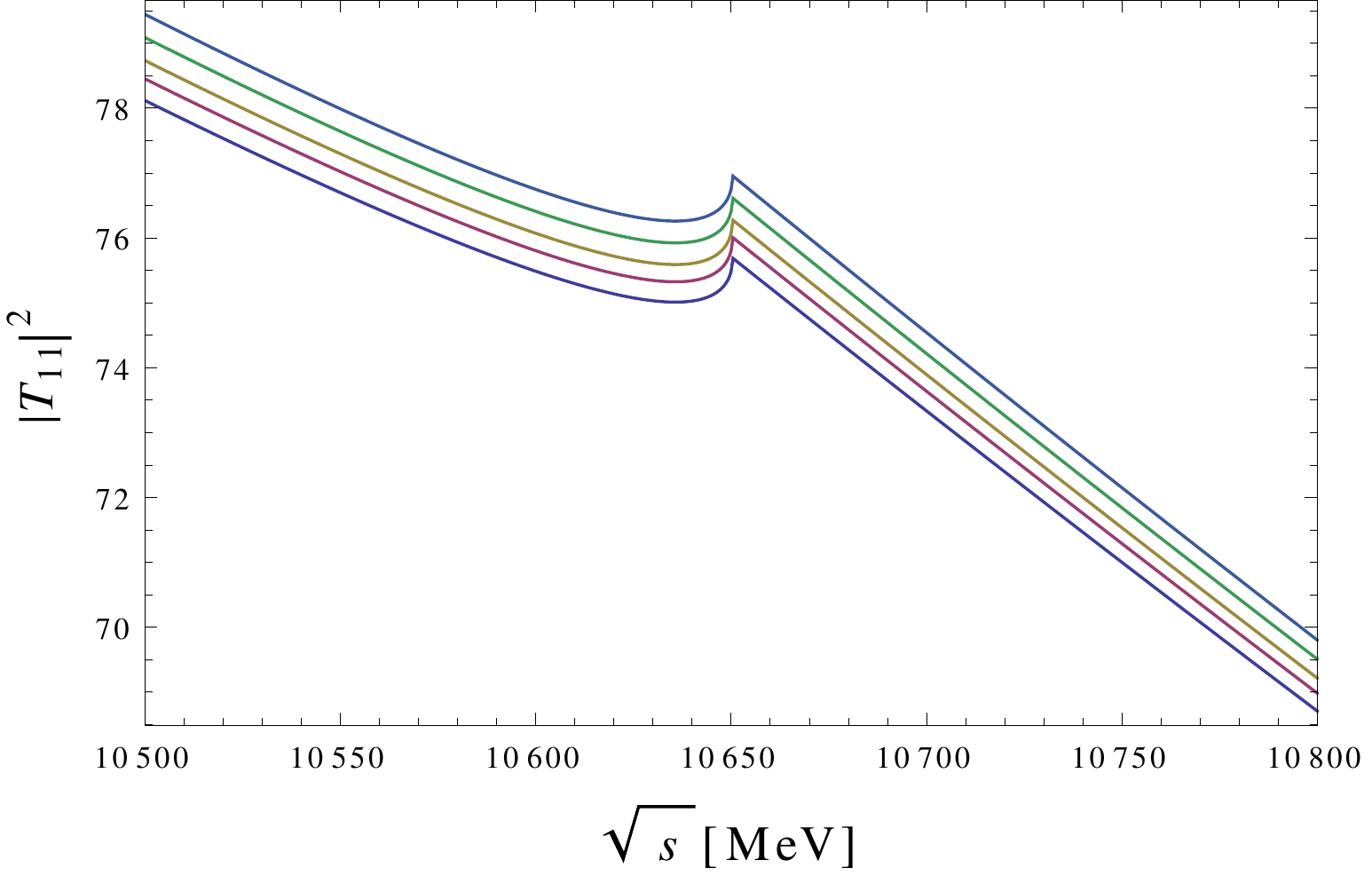}
\caption{$|T_{11}|^{2}$ as a function of the $\sqrt{s}$ center of mass energy when only the $B^*\bar{B}^*$ channel is considered ($J=1$ case). Each curve is related to the cut off values $q_{max}$ equal to $700,\,800,\,900,\,1000$ and $1100$ MeV. The peak moves from bottom to top as the cut off increases.}
\label{fig:TBsBsvec}
\end{figure}

In Fig. \ref{fig:TBsBsvec} we show the plot for $|T_{11}|^2$ as a function of the center of mass energy of the system. Note that in this case, we also have a peak about $10650$ MeV, which is just the threshold mass of the $B^*\bar{B}^*$ channel. Again, we see essentially a cusp in the amplitude which does not correspond to a bound state. The situation is similar if we increase the value of $\tilde{t}_{B^{*}\bar{B}^{*}\rightarrow B^{*}\bar{B}^{*}}$ of a factor $1.5$ to account for possible uncertainties. The value of $|T_{11}|^{2}$ grows accordingly, but the cusp remains and its shape is like in Fig. \ref{fig:TBsBsvec}.
 

\section{Summary and Conclusion}


Using the local hidden gauge lagrangians, we have studied the $B\bar{B}^*$ and $B^*\bar{B}^*$ interactions for isospin $I=1$. We show that the exchange of a light meson is not allowed by OZI rule. For that reason we have investigated the contributions coming from heavy vector exchange and also due to the two pion exchange, interacting and noninteracting among themselves, in which the OZI restriction no longer holds. Unlike Refs. \cite{Aceti:2014kja,Aceti:2014uea}, the vector exchange potential is not the main source of the interactions here. In view of this, we consider the vector exchange potential corrected by a factor that takes into account the contributions of the others mesons exchange cases and, then we use it as the kernel of the Bethe-Salpeter equation in order to solve the transition matrix $T$. Looking for poles in the $T$ matrix, we tried to relate them with the $Z_b(10610)$ and $Z_b(10650)$ states reported by Belle collaboration. From our results, using a cut off value $q_{max}=700$ MeV, we found a bound state of $B\bar{B}^*$ with mass in the range $10587 - 10601$ MeV very close to the experimental mass of the $Z_b(10610)$ at $10608$ MeV. In the case of $B^*\bar{B}^*$ interaction, we found a cusp at $10650$ MeV for spin $J=0$ and $J=2$ cases. On the other hand, the spin $J=1$ case can be considered only in the one channel problem without taking into account the $\rho\,\Upsilon$ channel. In this case, again a cusp at $10650$ MeV appears in the $|T_{11}|^2$ as can be seen in Fig. \ref{fig:TBsBsvec} and was also pointed out in Ref. \cite{Bugg:2011jr,Swanson:2014tra}.

\section*{Acknowledgments}
This work is partly supported by the Spanish Ministerio de Economia y Competitividad and European FEDER funds under the contract number
FIS2011-28853-C02-01, and the Generalitat Valenciana in the program Prometeo II, 2014/068. We acknowledge the support of the European Community-Research Infrastructure Integrating Activity Study of Strongly Interacting Matter (acronym HadronPhysics3, Grant Agreement
n. 283286) under the Seventh Framework Programme of EU. J. M. Dias would like to thank the Brazilian funding agency FAPESP for the financial support.


\begin{thebibliography}{99}


\bibitem{Choi:2003ue} 
  S.~K.~Choi {\it et al.}  [Belle Collaboration],
  Phys.\ Rev.\ Lett.\  {\bf 91}, 262001 (2003)
  [hep-ex/0309032].

\bibitem{x3872}
 V.~M.~Abazov et al. (D0 Collaboration), Phys. Rev. Lett. {\bf 93}, 162002 (2004); D. Acosta et al. (CDF Collaboration), Phys. Rev. Lett. {\bf 93}, 072001 (2004); B. Aubert et al. (BABAR Collaboration), Phys. Rev. D {\bf 71}, 071103 (2005).
\bibitem{Zhu:2007wz}
  S.~L.~Zhu,  Int.\ J.\ Mod.\ Phys.\ E {\bf 17}, 283 (2008) [hep-ph/0703225].  
\bibitem{Nielsen:2009uh}
  M.~Nielsen, F.~S.~Navarra and S.~H.~Lee,
  Phys.\ Rept.\  {\bf 497}, 41 (2010)  [arXiv:0911.1958 [hep-ph]].  

\bibitem{Ali:2011vy}
A.~Ali,
PoS BEAUTY {\bf 2011}, 002 (2011).

\bibitem{Brambilla:2010cs} 
  N.~Brambilla, S.~Eidelman, B.~K.~Heltsley, R.~Vogt, G.~T.~Bodwin, E.~Eichten, A.~D.~Frawley and A.~B.~Meyer {\it et al.},
  Eur.\ Phys.\ J.\ C {\bf 71}, 1534 (2011).

\bibitem{Swanson:2006st} 
  E.~S.~Swanson,
  Phys.\ Rept.\  {\bf 429}, 243 (2006).

\bibitem{Olsen:2009gi}
  S.~L.~Olsen,
  Nucl.\ Phys.\  {\bf A827}, 53C-60C (2009) [arXiv:0901.2371].
  
\bibitem{Chen:2011}
W.~Chen and S.~L.~Zhu,
Phys.\ Rev.\ D {\bf 83}, 034010 (2011).

\bibitem{Belle:2011aa}
  A.~Bondar {\it et al.}  [Belle Collaboration],
  Phys.\ Rev.\ Lett.\  {\bf 108}, 122001 (2012)
  [arXiv:1110.2251 [hep-ex]].

\bibitem{Adachi:2012im} 
  I.~Adachi {\it et al.}  [Belle Collaboration],
  arXiv:1207.4345 [hep-ex].

\bibitem{Bondar:2011ev} 
  A.~E.~Bondar, A.~Garmash, A.~I.~Milstein, R.~Mizuk and M.~B.~Voloshin,
  Phys.\ Rev.\ D {\bf 84}, 054010 (2011)
  [arXiv:1105.4473 [hep-ph]].


\bibitem{Bugg:2011jr} 
  D.~V.~Bugg,
  Europhys.\ Lett.\  {\bf 96}, 11002 (2011)
  [arXiv:1105.5492 [hep-ph]].
  
\bibitem{Swanson:2014tra}
  E.~S.~Swanson,
  arXiv:1409.3291 [hep-ph].

\bibitem{Danilkin:2011sh} 
  I.~V.~Danilkin, V.~D.~Orlovsky and Y.~.A.~Simonov,
  Phys.\ Rev.\ D {\bf 85}, 034012 (2012)
  [arXiv:1106.1552 [hep-ph]].

\bibitem{Cui:2011fj} 
  C.~-Y.~Cui, Y.~-L.~Liu and M.~-Q.~Huang,
  Phys.\ Rev.\ D {\bf 85}, 074014 (2012)
  [arXiv:1107.1343 [hep-ph]].

\bibitem{Guo:2011gu} 
  T.~Guo, L.~Cao, M.~-Z.~Zhou and H.~Chen,
  arXiv:1106.2284 [hep-ph].


\bibitem{Sun:2011uh} 
  Z.~-F.~Sun, J.~He, X.~Liu, Z.~-G.~Luo and S.~-L.~Zhu,
  Phys.\ Rev.\ D {\bf 84}, 054002 (2011)
  [arXiv:1106.2968 [hep-ph]].


\bibitem{Cleven:2011gp} 
  M.~Cleven, F.~-K.~Guo, C.~Hanhart and U.~-G.~Meissner,
  Eur.\ Phys.\ J.\ A {\bf 47}, 120 (2011)
  [arXiv:1107.0254 [hep-ph]].


\bibitem{Voloshin:2011qa} 
  M.~B.~Voloshin,
  Phys.\ Rev.\ D {\bf 84}, 031502 (2011)
  [arXiv:1105.5829 [hep-ph]].


\bibitem{Zhang:2011jja} 
  J.~-R.~Zhang, M.~Zhong and M.~-Q.~Huang,
  Phys.\ Lett.\ B {\bf 704}, 312 (2011)
  [arXiv:1105.5472 [hep-ph]].

\bibitem{Ali:2011ug} 
  A.~Ali, C.~Hambrock and W.~Wang,
  Phys.\ Rev.\ D {\bf 85}, 054011 (2012)
  [arXiv:1110.1333 [hep-ph]].

\bibitem{Yang:2011rp} 
  Y.~Yang, J.~Ping, C.~Deng and H.~-S.~Zong,
  J.\ Phys.\ G {\bf 39}, 105001 (2012)
  [arXiv:1105.5935 [hep-ph]].

\bibitem{Chen:2011zv} 
  D.~-Y.~Chen, X.~Liu and S.~-L.~Zhu,
  Phys.\ Rev.\ D {\bf 84}, 074016 (2011)
  [arXiv:1105.5193 [hep-ph]].


\bibitem{Nieves:2011zz} 
  J.~Nieves and M.~P.~Valderrama,
  Phys.\ Rev.\ D {\bf 84}, 056015 (2011)
  [arXiv:1106.0600 [hep-ph]].


\bibitem{Mehen:2011yh} 
  T.~Mehen and J.~W.~Powell,
  Phys.\ Rev.\ D {\bf 84}, 114013 (2011)
  [arXiv:1109.3479 [hep-ph]].

\bibitem{Chen:2011pv} 
  D.~-Y.~Chen and X.~Liu,
  Phys.\ Rev.\ D {\bf 84}, 094003 (2011)
  [arXiv:1106.3798 [hep-ph]].


\bibitem{Cleven:2013sq} 
  M.~Cleven, Q.~Wang, F.~-K.~Guo, C.~Hanhart, U.~-G.~Meissner and Q.~Zhao,
  Phys.\ Rev.\ D {\bf 87}, 074006 (2013)
  [arXiv:1301.6461 [hep-ph]].


\bibitem{Dong:2012hc} 
  Y.~Dong, A.~Faessler, T.~Gutsche and V.~E.~Lyubovitskij,
  J.\ Phys.\ G {\bf 40}, 015002 (2013)
  [arXiv:1203.1894 [hep-ph]].


\bibitem{Braaten:2013boa} 
  E.~Braaten,
  arXiv:1305.6905 [hep-ph].

\bibitem{Valderrama:2012jv} 
  M.~P.~Valderrama,
  Phys.\ Rev.\ D {\bf 85}, 114037 (2012)
  [arXiv:1204.2400 [hep-ph]].

\bibitem{Ke:2012gm} 
  H.~-W.~Ke, X.~-Q.~Li, Y.~-L.~Shi, G.~-L.~Wang and X.~-H.~Yuan,
  JHEP {\bf 1204}, 056 (2012)
  [arXiv:1202.2178 [hep-ph]].

\bibitem{Ohkoda:2012rj} 
  S.~.Ohkoda, Y.~.Yamaguchi, S.~.Yasui and A.~.Hosaka,
  Phys.\ Rev.\ D {\bf 86}, 117502 (2012)
  [arXiv:1210.3170 [hep-ph]].


\bibitem{Li:2012wf} 
  M.~T.~Li, W.~L.~Wang, Y.~B.~Dong and Z.~Y.~Zhang,
  J.\ Phys.\ G {\bf 40}, 015003 (2013)
  [arXiv:1204.3959 [hep-ph]].

\bibitem{Wang:2013zra} 
  Z.~-G.~Wang and T.~Huang,
  arXiv:1312.2652 [hep-ph].

\bibitem{Guo:2013gka} 
  F.~-K.~Guo, C.~Hidalgo-Duque, J.~Nieves and M.~P.~Valderrama,
  arXiv:1309.3865 [hep-ph].

\bibitem{Zhi:2011liu}
Zhi-Feng Sun, Jun He, and Xiang Liu,
Phys.\ Rev.\ D {\bf 84}, 054002 (2011).

\bibitem{Aceti:2014uea} 
  F.~Aceti, M.~Bayar, E.~Oset, A.~M.~Torres, K.~P.~Khemchandani, F.~S.~Navarra and M.~Nielsen,
  Phys.\ Rev.\ D {\bf 90}, 016003 (2014).
  
\bibitem{Aceti:2014kja} 
  F.~Aceti, M.~Bayar, J.~M.~Dias and E.~Oset,
  Eur.\ Phys.\ J.\ A {\bf 50}, 103 (2014).
  
\bibitem{xiaoliang} 
  W.~H.~Liang, C.~W.~Xiao and E.~Oset,
Phys.\ Rev.\ D {\bf 89}, 054023 (2014).
  
\bibitem{xiaojuan}
  C.~W.~Xiao, J.~Nieves and E.~Oset,
  Phys.\ Rev.\ D {\bf 88}, 056012 (2013)
  [arXiv:1304.5368 [hep-ph]].

\bibitem{hidden1}
  M.~Bando, T.~Kugo, S.~Uehara, K.~Yamawaki and T.~Yanagida,
  Phys.\ Rev.\ Lett.\  {\bf 54}, 1215 (1985).

\bibitem{hidden2}
  M.~Bando, T.~Kugo and K.~Yamawaki,
  Phys.\ Rept.\  {\bf 164}, 217 (1988).

\bibitem{hidden3}
  M.~Harada and K.~Yamawaki,
  Phys.\ Rept.\  {\bf 381}, 1 (2003)

\bibitem{raquelxyz}
R.~Molina and E.~Oset,
Phys.\ Rev.\ D {\bf 80}, 114013 (2009).

\bibitem{gamphi3770}
  D.~Gamermann, E.~Oset and B.~S.~Zou,
  Eur.\ Phys.\ J.\  A {\bf 41}, 85 (2009)  

\bibitem{daniel} 
  D.~Gamermann and E.~Oset,
  Eur.\ Phys.\ J.\ A {\bf 33}, 119 (2007).

\bibitem{luisaxial} 
  L.~Roca, E.~Oset and J.~Singh,
  Phys.\ Rev.\ D {\bf 72}, 014002 (2005).
  
\bibitem{daniwf}
  D.~Gamermann, J.~Nieves, E.~Oset and E.~Ruiz Arriola,
  Phys.\ Rev.\ D {\bf 81}, 014029 (2010)
  [arXiv:0911.4407 [hep-ph]].

\bibitem{toki}
  E.~Oset, H.~Toki, M.~Mizobe and T.~T.~Takahashi,
  Prog.\ Theor.\ Phys.\  {\bf 103}, 351 (2000).

\bibitem{Olsen:2012zz}
S.~L.~Olsen,
Prog.\ Theor.\ Phys.\ Suppl.\ {\bf 193}, 38 (2012).

\bibitem{yama}
  J.~Yamagata-Sekihara, J.~Nieves and E.~Oset,
  Phys.\ Rev.\ D {\bf 83}, 014003 (2011)
  [arXiv:1007.3923 [hep-ph]].
  
\bibitem{Rubin:2004cq}
  P.~Rubin {\it et al.}  [CLEO Collaboration],
  Phys.\ Rev.\ Lett.\  {\bf 93}, 111801 (2004)
  [hep-ex/0405011].

\bibitem{Oller:1998hw}
  J.~A.~Oller, E.~Oset and J.~R.~Pelaez,
  Phys.\ Rev.\ D {\bf 59}, 074001 (1999)
  [Erratum-ibid.\ D {\bf 60}, 099906 (1999)]
  [Erratum-ibid.\ D {\bf 75}, 099903 (2007)]
  [hep-ph/9804209].
  
\end{thebibliography}
\end{document}